\documentclass{article}
\begin{document}

\thispagestyle{plain}
\def\upcirc#1{\vbox{\ialign{##cr
$\circ\!$cr \noalign{\kern-0.1 pt\nointerlineskip}
$\hfil\displaystyle{#1}\hfil$cr}}}
\def\X{\bf X}
\def\thorn{\hbox{\rm I}\kern-0.32em\raise0.35ex\hbox{\it o}}
\def\edth{\hbox{$\partial$\kern-0.25em\raise0.6ex\hbox{\rm\char'40}}}
\def\edthbar{\overline{\edth}}
\def\taub{\overline{\tau}}
\def\obar{\overline{o}}
\def\iotabar{\overline{\iota}}
\def\thornprime{\thorn^\prime}
\def\edthprime{\edth^\prime}
\def\i{\iota}
\def\P{{\bf \Psi}}
\def\Thorn{{\bf \thorn}}
\def\Fi{{\bf \Phi}}
\def\bfi{\bf \phi}
\def\bnab{{\bf \nabla}}
\def\L{{\bf \Lambda}}
\def\bxi{{\bf \Xi}}
\def\bep{{\bf \epsilon}}
\def\bdelta{\mbox{\boldmath $\delta$}}
\def\bDelta{{\bf \Delta}}
\def\thornp{{\thorn^\prime}}
\def\edthp{{\edth^\prime}}
\def\bfeta{{\bf \eta}}
\def\pmb#1{\setbox0=\hbox{#1} \kern-.025em\copy0\kern-\wd0
\kern.05em\copy0\kern-\wd0 \kern-.025em\raise.0233em\box0 }
\def\bedth{{\pmb{\edth}}}
\def\bthorn{{\pmb \thorn}}
\def\thornp{{\thorn^\prime}}
\def\edthp{{\edth^\prime}}
\def\bfeta{{\bf \eta}}
\def\bthornp{{\bthorn^\prime}}
\def\bedthp{{\bedth^\prime}}
\def\alemf{{\nabla^{f\plf}\nabla_{f\plf}}}
\def\Ib{\overline{\bf I}}
\def\I{\bf I}
\def\ei{{\rm i}}
\def\Ph{\nthorn}
\def\D{\nedth}
\parindent=0pt
\def\pmb#1{\setbox0=\hbox{#1}  \kern-.025em\copy0\kern-\wd0
 \kern.05em\copy0\kern-\wd0
  \kern-0.025em\raise.0433em\box0 }
\def\half{{\scriptstyle {1 \over 2}}}
\def\third{{\scriptstyle {1 \over 3}}}
\def\quarter{{\scriptstyle {1 \over 4}}}
\def\o{{\pmb o}}
\def\i{{\pmb{$\iota$}}}
\let\gt=\mapsto
\let\la=\lambda
\def\a{\overline a}
\def\daa{\nabla_{AA'}}
\def\ob{{\overline\o}}
\def\ib{{\overline\i}}
\def\T{{\bf T}}
\def\bp{{\bf p}}
\def\bq{{\bf q}}
\def\0{\pmb 0}
\def\1{\pmb 1}
\def\2{\pmb 2}
\def\3{\pmb 3}
\def\bPhi{{\pmb{$\Phi$}}}
\def\bPsi{{\pmb{$\Psi$}}}
\def\>{\phantom{A}}
\def\sym{{\sum_{sym}}}
\def\thorn{\hbox{\rm I}\kern-0.32em\raise0.35ex\hbox{\it o}}
\def\edth{\hbox{$\partial$\kern-0.25em\raise0.6ex\hbox{\rm\char'40}}}
\def\edthbar{\overline{\edth}}
\def\nedth{{\pmb\edth}}
\def\nedthp{{\pmb\edth'}}
\def\thornp{\thorn'}
\def\nthorn{{\pmb\thorn}}
\def\nthornp{{\pmb\thorn'}}
\def\edthp{\edth'}
\def\a{\alpha}
\def\b{\beta}
\def\g{\gamma}
\def\d{\delta}
\def\cd{{\cal D}}
\def\boeta{{\pmb{$\eta$}}}
\def\blambda{{\pmb{$\lambda$}}}
\def\etad{\boeta_{C_1\dots C_N C'_1\dots C'_{N'}}}
\def\p{{\bf p}}
\def\q{{\bf q}}
\def\K{{\bf K}}
\def\R{{\bf R}}
\def\S{{\bf S}}
\def\T{{\bf T}}
\def\I{{\bf I}}
\def\la{\lambda}
\def\lab{\overline\lambda}

\title{Type O pure radiation metrics with a cosmological constant.}
\author{S. Brian Edgar
\\ Department of Mathematics,
 \\ Link\"{o}pings universitet\\ Link\"{o}ping,\\
Sweden S-581 83\\
 email: bredg@mai.liu.se
\and M.P. Machado Ramos
\\ Departamento de Matem\' atica
\\ para a Ci\^encia e Tecnologia,
\\ Azur\'em 4800-058 Guimar$\tilde{\hbox {a}}$es,
\\
Universidade do Minho,
\\ Portugal\\
 email: mpr@mct.uninho.pt}

\maketitle

\section*{Abstract.}

 In this paper we complete the integration of the conformally flat pure radiation spacetimes with a non-zero cosmological constant $\Lambda$, and $\tau \ne 0$, by considering the case $\Lambda +\tau\bar\tau \ne 0$. This is a further demonstration of the power and suitability of the generalised invariant formalism (GIF) for spacetimes where only one null direction is  picked out by the Riemann tensor.  For these spacetimes, the GIF picks out a second null direction, (from the second derivative of the Riemann tensor) and once this spinor has been identified the calculations are transferred to the simpler GHP formalism, where the tetrad and metric are determined. The whole class of conformally flat pure radiation spacetimes with a non-zero cosmological constant (those found in this paper, together with those found earlier for the case $\Lambda +\tau\bar\tau = 0$) have  a rich variety of subclasses with zero, one, two, three, four or five Killing vectors.

\

{\bf PACS numbers:} \ 0420,\ 1127

\

\section{Introduction}
\subsection{Integration in GHP formalism}

The method of integration within the Geroch-Held-Penrose (GHP) formalism
using GHP operators \cite{ghp} pioneered by Held \cite{held1}, \cite{held2}
and developed by Edgar and Ludwig \cite{edGHP}, \cite{edlud1}, \cite{edlud2}%
, \cite{edlud3} has been shown to be particularly useful and efficient in
spacetimes where two null directions are picked out by the geometry. The
'optimal situation' is when the GHP formalism generates internally a set of tables
involving the first derivative GHP operators for each of four real
zero-weighted \textit{intrinsic} scalars ('coordinate candidates') and a
table for one complex weighted ($p\neq 0\neq q)$ \textit{intrinsic} scalar
(which describes the spin ($(p-q)/2$) and boost ($(p+q)/2$) gauge). An
important ingredient within this method is the repeated application of the
GHP commutator equations; in particular, it is essential that these
commutators be applied to these five scalars in order to extract all the
information residing in the GHP commutators \cite{edGHP}. Held's original hope
was that these five tables, by themselves,  would be complete and involutive: in general, we
now know that they will not be, since  the manipulations --- especially the
application of the commutators to the coordinate candidates --- will generate
additional scalars and their associated tables; however taking  all these
tables together ensures a complete and involutive system. The development
and applications of this method can be found in \cite{held1}, \cite{held2},
\cite{edGHP}, \cite{edlud1}, \cite{edlud3},  \cite{kerr}, \cite
{carm1}, \cite{carm2}.

One of the intriguing aspects of this operator method in the GHP formalism
is that the 'optimal situation' --- where the calculations for constructing
a metric are, in principle, simplest --- is for spaces lacking Killing
vectors, and lacking spin and boost isotropy freedom, \cite{edlud3}. (On
the other hand, it is well known that the NP tetrad formalism \cite{np}  has
been particularly useful in investigating spaces with isotropy freedom
and/or Killing vectors.) From the theory and structure of the GHP formalism,
it follows that the presence of the full quota of four
{\it intrinsic} coordinate candidates is directly linked to the absence of Killing
vectors, while the presence of a complex weighted {\it intrinsic} scalar is
directly linked to the absence of spin and boost isotropy freedom; in spaces
with no Killing vectors there will be four  {\it intrinsic} coordinates within the formalism, and in spaces with no spin and boost
isotropy freedom there will be one complex {\it intrinsic} weighted scalar. In the
situation where less than four zero-weighted \textit{intrinsic} scalars are supplied
directly by the GHP formalism, it is then necessary to introduce
replacements for the absent coordinate candidate(s) indirectly --- each via
a table which is complementary to the complete explicit set of GHP equations, being
entirely consistent with them; in particular the table(s) for the \textit{%
complementary} zero-weighted scalar(s) ({\it complementary coordinate candidate(s)}) must be consistent with the GHP
commutators. In an analogous manner, in the situation where a complex
weighted \textit{intrinsic} scalar is not supplied directly by the GHP formalism, it
is then necessary to introduce a replacement for this absent complex scalar (or
absent part of the complex scalar) indirectly --- via a table which is
complementary to the complete explicit set of GHP equations; the table for the \textit{%
complementary weighted scalar} also must be consistent with all the GHP
equations including the GHP commutators. In addition, all the new tables must be consistent with each other. It is emphasised
that it is essential that the full quota of five scalars be obtained (by
supplementing intrinsic ones with complementary ones, where necessary), and
that the GHP commutators be applied to each, so as to ensure that \textit{all} the
information in the GHP commutators is extracted.

Using this integration procedure in the GHP formalism, in addition to
 applications to spacetimes without Killing vectors, there have been
applications to spaces with Killing vectors (mostly to spaces with one
Killing vector) \cite{edGHP}, \cite{edlud1}, \cite{kerr}, and
a technique has emerged for introducing the tables for the complementary
coordinate candidates. For example, in the case of a spacetime with only
three intrinsic coordinate candidates $x_{1},x_{2},x_{3}$ and requiring a
complementary coordinate candidate $\tilde{x}_{4}$, the idea is to make use
of a related 'generic' spacetime (where it exists) with the analogous three
intrinsic coordinates candidates $x_{1},x_{2},x_{3}$ plus a fourth intrinsic
coordinate $x_{4}$; then to introduce the complementary coordinate candidate
$\tilde{x}_{4}$ via a `copy' of the table for the coordinate candidate $x_{4}$%
, but in addition `freeing' this complementary coordinate candidate $\tilde{x}%
_{4}$ from any {\it direct} links which $x_{4}$ had with the remaining explicit elements of
the GHP formalism.

It has been shown that spacetimes constructed by this method not only can be
easily analysed for their Killing vector stucture, but the explicit form of
the Killing vectors (and even the homothetic  vectors)
can be found in a comparatively easy manner \cite{edlud3}.

\subsection{Integration in GIF}

The generalised invariant formalism (GIF) of Machado Ramos and
Vickers \cite{maria1}, \cite{maria2}, \cite{maria3} generalises the GHP
formalism by  building the null rotation freedom of the second
null direction into the formalism, which means that the GIF is built around
only one spinor $\mathbf{o}_{A}$. However, the formal set of equations for
the GIF is considerably more complicated than for the GHP formalism,
involving \textit{spinor} differential operators; this means that calculations ---
especially involving the GIF commutators --- are much more involved.
Nevertheless, an analogous integration method \cite{edvic}, \cite{edram1}, \cite{edram2}
using operators of the GIF has been developed. Once again, the 'optimal
situation' is when the formalism generates a set of tables involving
first derivative  operators for each of four real zero-weighted \textit{%
intrinsic} scalars and  for one complex weighted \textit{intrinsic}
scalar: but in addition, we need to generate a table for a second \textit{intrinsic} spinor $%
\mathbf{I}_{A}$ (which is not parallel to the first spinor $\mathbf{o}_{A}$).
As with the GHP procedure, the integration technique in the GIF relies heavily on repeated applications of the
commutators, and it is essential that the GIF commutators be applied to the full
quota of five scalars, and also to the second spinor $\mathbf{I}_{A}$.
Hence, if we are investigating a less than 'optimal situation' which fails to
generate the full quota of  \textit{intrinsic} scalars, then \textit{%
complementary} scalars will need to be introduced indirectly via tables as
in the GHP method, while if we cannot generate a second unique \textit{%
intrinsic} spinor, a \textit{complementary} spinor will also need to be
introduced indirectly via a table.

\smallskip The very first investigation using the GIF integration method was
for the class of conformally flat pure radiation spacetimes (with zero
cosmological constant) \cite{edvic}. The pure radiation component of the
Ricci tensor immediately picks out one null direction $\mathbf{o}_{A}$, and
the generic class of these spacetimes admits no Killing vectors; so these
spaces were particularly well suited for investigation by this approach in
the GIF \cite{edvic}. A second intrinsic spinor $\mathbf{I}_{A}$ was
obtained after a little manipulation in the GIF: the optimal
situation was achieved in the generic case --- with no Killing vectors; and
all four intrinsic zero-weighted scalars, together with an intrinsic complex
weighted scalar, were generated inernally within  the GIF. The GIF commutators were applied to the second spinor and to the five scalars.  As a consequence, a complete and
involutive set of tables was obtained in the GIF.  For the non-generic case, which required one complementary coordidate
candidate --- corresponding to the presence of one Killing vector ---  a parallel calculation
in the GIF was carried out, with a table for the complementary coordinate
'copied' from the generic case. Finally, for both cases, by identifying
the spinor $\mathbf{I}_{A}$ with the second dyad spinor ${{\setbox0=%
\hbox{$\iota$}\kern-.025em\copy0\kern-\wd0\kern.05em\copy0\kern-\wd0\kern%
-0.025em\raise.0433em\box0}}_{A}$ of the GHP formalism, the investigations
were transfered into, and completed in, the GHP formalism; the final step involved deducing the tetrad vectors from the GHP tables for the four coordinate candidates, and hence the metric.
In \cite{edvic}, the part of the investigation in the GIF which established the complete and
involutive set of GIF tables was the most complicated; in particular because
of the repeated use of the complicated GIF commutators.

In fact, with the benefit of hindsight, it is now clear that it was not necessary to carry out \textit{all} of these
GIF calculations in \cite{edvic}: the crucial step in the GIF was to generate explicitly
this second intrinsic spinor $\mathbf{I}_{A}$ from within the GIF formalism.
As soon as this spinor was found and the GIF commutators applied to it, then $%
\mathbf{I}_{A}$ could have been identified as the second spinor ${{\setbox0=%
\hbox{$\iota$}\kern-.025em\copy0\kern-\wd0\kern.05em\copy0\kern-\wd0\kern%
-0.025em\raise.0433em\box0}}_{A}$ in the dyad for the GHP formalism; at this stage,
the investigation could have been \textit{immediately} transferred to the
GHP formalism. Had this earlier transfer been made, the latter part of the
complicated GIF calculations in \cite{edvic} could have been replaced with
simpler GHP calculations.

On the other hand, it is emphasised that when we fail to obtain a unique
\textit{intrinsic} second spinor we cannot take this
short cut to the GHP formalism. The absence of a second unique \textit{intrinsic} spinor is linked to the
presence of null isotropy freedom, and we have recently considered such a
situation --- a subclass of conformally flat pure radiation spacetimes with
a cosmological constant --- and demonstrated how such a problem is solved by
the GIF method \cite{edram2}.

\subsection{Outline of paper}

In this paper we wish to develop the operator method further and more
efficiently in the GIF formalism, by investigating the remainder of the conformally flat pure radiation spacetimes with a cosmological constant. First of all, we wish to illustrate the point emphasised above:
when we are able to determine a second unique intrinsic spinor $\mathbf{I}_{A}$ in
the GIF formalism we will demonstrate how to transfer to the  simpler GHP formalism, and thus reduce the amount of calculations. In
order to do this we will need to obtain the GHP commutators, which are
crucial to simplifying our analysis; we shall demonstrate how to do this in
a simple manner. Secondly, by investigating a class of spacetimes with a
richer Killing vector structure than the cases looked at before, we will
learn more about how to introduce tables for complementary coordinate
candidates in the GIF, by modifying the techniques which have been outlined above for the GHP formalism.

The approach being adopted in this paper, (as was also adopted in \cite%
{edvic} and \cite
{edram2}), is to attempt to generate the complex scalar and as many coordinates as possible \textit{%
intrinsically}, i.e., directly from elements of the GIF/GHP formalisms; it is only
when it is clear that no more intrinsic coordinate candidates are available  that we introduce {\it complementary}
coordinate candidates by their respective tables. An advantage of this approach is
that subsequently we can easily interpret the Killing vector structure and
Karlhede classification from this version of the metric, which will involve  the
maximum number of 'good' coordinates. An alternative approach would be to
introduce complementary coordinates, even when we suspect that there
may exist intrinsic coordinates which we have not exploited; such an
approach may be advantageous when the intrinsic choices lead to very
complicated calculations. This alternative approach was carried out in our
investigation of Petrov type N pure radiation spacetimes \cite{edram1};
to avoid misunderstandings, we emphasise the different approach in that
paper to this one.

So now we will further develop the GIF operator method by generalising the
earlier derivation \cite{edvic} of the metric for conformally flat pure
radiation spaces to include the case of a non-zero cosmological constant. In
\cite{edram2} we looked at the subclass of these spacetimes for which we
were unable to find a second unique \textit{intrinsic} spinor due to the
presence of one degree of null isotropy freedom; in this paper we look at
the other subclass where there is no null isotropy freedom, and a second
unique \textit{intrinsic} spinor  $\mathbf{I}_{A}$ is quickly generated
within the GIF. This means that we can quickly transfer to the GHP formalism
and so minimise the calculations.
These spacetimes will illustrate further refinements of our
method, and they will also be shown o have a richer Killing vector structure.

More details of the philosophy and techniques of the GIF
operator integration procedure has been given in \cite{edvic}, \cite
{edram1}, \cite{edram2} so we will not repeat these discussions in this paper, but rather
we will only summarise the relevant parts of the GIF  which are
needed in this paper. In Section \ref{gif} we describe the differential
operators, and the equations for the class of spaces under consideration are
given in Section \ref{equations}. In Section  \ref{transfer} we discuss the
principles of the early transfer from the GIF to the GHP formalism, and how to obtain the GHP commutator equations.

In the beginning of Section \ref{intne}, in order to make comparison easy,
we carry through the integration procedure initially keeping close to the
pattern of the calculations in \cite{edvic}, obtaining a table for the
crucial second spinor $\mathbf{I}_{A}$. As
soon as we  obtain this second unique \textit{intrinsic} spinor $\mathbf{I
}_{A}$, and extract additional information by applying the GIF commutators to it, we translate all the results  into the GHP formalism;  in the latter part of this
section we show, by a straightforward relabelling and rearranging of some of
the coordinate candidates and unknown functions, that their GHP tables can
be put into a much simpler form, so that we can more easily complete the
application of the commutators to these tables. Finally, from these tables,
we write down the tetrad and the metric explicitly.

The procedure in Section \ref{intne} is dependent on the condition that the
four zero-weighted scalars, to which we assign the role of coordinate
candidates, are functionally independent and hence can play the role of
coordinates; indeed, if we make the assumption that none of these scalars
are constants, then a check of the determinant formed from their four tables
shows that all four scalars are in fact  functionally independent. On the
other hand, it is found that although three of the four coordinate
candidates cannot be
constant, the other one may be;  in addition, we make some other assumptions in our calculations which exclude some other special cases. Hence the tetrad and metric obtained in Section \ref%
{intne} are not the most general that can be obtained for this class of
spacetimes.

In Section \ref{special}, we extend our approach to
include one of the special cases which were excluded in the analysis in Section \ref{intne}.

In Section \ref{all} we consider
the remaining special case and discuss in more detail the introduction and role of {\it complementary} coordinates, and how to {\it copy} their tables; also in that section we put together all the subclasses and present the most general form for the metric.
 In Section 8 we summarise the methods and results of this paper together with
those of \cite{edram2}.

\section{GIF }

\label{gif}

A full explanation of the formalism is given in \cite{maria1}, \cite{maria2}, \cite{maria3}.
For the purpose of this paper, the summaries given in \cite{edvic} and especially \cite{edram2} are sufficient.

In this subsection, we will list only those equations to which we will make direct reference.
The GIF differential operators ${\setbox0=\hbox{\thorn} \kern-.025em\copy0%
\kern-\wd0 \kern.05em\copy0\kern-\wd0 \kern-0.025em\raise.0433em\box0 } $, ${%
\setbox0=\hbox{\edth} \kern-.025em\copy0\kern-\wd0 \kern.05em\copy0\kern-\wd%
0 \kern-0.025em\raise.0433em\box0 }$, ${\setbox0=\hbox{\thorn} \kern-.025em%
\copy0\kern-\wd0 \kern.05em\copy0\kern-\wd0 \kern-0.025em\raise.0433em\box0
^{\prime}}$ and ${\setbox0=\hbox{\edth} \kern-.025em\copy0\kern-\wd0 \kern%
.05em\copy0\kern-\wd0 \kern-0.025em\raise.0433em\box0 ^{\prime}}$ act
on properly weighted symmetric spinors to produce symmetric spinors of
different valence and weight.
Although the definition of the differential operators appears quite
complicated, the fact that they take symmetric spinors to symmetric spinors
means that one can write down the equations in a more compact and index free
notation. In this compacted notation we have the following useful identities for scalars of weight $\{p,q\}$,
\begin{eqnarray}
({\setbox0=\hbox{\thorn}\kern-.025em\copy0\kern-\wd0\kern.05em\copy0\kern-\wd%
0\kern-0.025em\raise.0433em\box0^{\prime }}\eta )\cdot {\overline{\setbox0=%
\hbox{o}\kern-.025em\copy0\kern-\wd0\kern.05em\copy0\kern-\wd0\kern-0.025em%
\raise.0433em\box0}}={\scriptstyle{\frac{1}{2}}}\{({\setbox0=\hbox{\edth}%
\kern-.025em\copy0\kern-\wd0\kern.05em\copy0\kern-\wd0\kern-0.025em\raise%
.0433em\box0^{\prime }}\eta )-q\overline{\mathbf{T}}\eta \}\label{thornp.ob}
\end{eqnarray}%
\begin{eqnarray}
({\setbox0=\hbox{\thorn}\kern-.025em\copy0\kern-\wd0\kern.05em\copy0\kern-\wd%
0\kern-0.025em\raise.0433em\box0^{\prime }}\eta )\cdot {\setbox0=\hbox{o}%
\kern-.025em\copy0\kern-\wd0\kern.05em\copy0\kern-\wd0\kern-0.025em\raise%
.0433em\box0}={\scriptstyle{\frac{1}{2}}}\{({\setbox0=\hbox{\edth}\kern%
-.025em\copy0\kern-\wd0\kern.05em\copy0\kern-\wd0\kern-0.025em\raise.0433em%
\box0}\eta )-p\mathbf{T}\eta \}
\label{thornp.o}
\end{eqnarray}%
\begin{eqnarray}
({\setbox0=\hbox{\edth}\kern-.025em\copy0\kern-\wd0\kern.05em\copy0\kern-\wd0%
\kern-0.025em\raise.0433em\box0^{\prime }}\eta )\cdot {\setbox0=\hbox{o}\kern%
-.025em\copy0\kern-\wd0\kern.05em\copy0\kern-\wd0\kern-0.025em\raise.0433em%
\box0}={\scriptstyle{\frac{1}{2}}}\{({\setbox0=\hbox{\thorn}\kern-.025em\copy%
0\kern-\wd0\kern.05em\copy0\kern-\wd0\kern-0.025em\raise.0433em\box0}\eta )-p%
\mathbf{R}\eta \}
\label{edthp.o}
\end{eqnarray}
\begin{eqnarray}
({\setbox0=\hbox{\edth}\kern-.025em\copy0\kern-\wd0\kern.05em\copy0\kern-\wd0%
\kern-0.025em\raise.0433em\box0}\eta )\cdot {\overline{\setbox0=\hbox{o}\kern%
-.025em\copy0\kern-\wd0\kern.05em\copy0\kern-\wd0\kern-0.025em\raise.0433em%
\box0}}={\scriptstyle{\frac{1}{2}}}\{({\setbox0=\hbox{\thorn}\kern-.025em%
\copy0\kern-\wd0\kern.05em\copy0\kern-\wd0\kern-0.025em\raise.0433em\box0}%
\eta )-q\overline{\mathbf{R}}\eta \}
\label{edth.ob}
\end{eqnarray}%
\begin{eqnarray}({\setbox0=\hbox{\thorn}\kern-.025em\copy0\kern-\wd0\kern.05em\copy0\kern-\wd%
0\kern-0.025em\raise.0433em\box0^{\prime }}\eta )\cdot {\setbox0=\hbox{o}%
\kern-.025em\copy0\kern-\wd0\kern.05em\copy0\kern-\wd0\kern-0.025em\raise%
.0433em\box0}\cdot {\overline{\setbox0=\hbox{o}\kern-.025em\copy0\kern-\wd0%
\kern.05em\copy0\kern-\wd0\kern-0.025em\raise.0433em\box0}}={\scriptstyle{%
\frac{1}{4}}}\{({\setbox0=\hbox{\thorn}\kern-.025em\copy0\kern-\wd0\kern.05em%
\copy0\kern-\wd0\kern-0.025em\raise.0433em\box0}\eta )-p\mathbf{R}\eta -q%
\overline{\mathbf{R}}\eta \}
\label{thornp.o.ob}
\end{eqnarray}%
For a spinor ${\setbox0=\hbox{$\eta$}\kern-.025em\copy0\kern-\wd0\kern.05em%
\copy0\kern-\wd0\kern-0.025em\raise.0433em\box0}$ the above contractions
become more complicated. For example for a valence (1,0)-spinor ${\setbox0=%
\hbox{$\eta$}\kern-.025em\copy0\kern-\wd0\kern.05em\copy0\kern-\wd0\kern%
-0.025em\raise.0433em\box0}_{A}$ of weight $\{\mathbf{p},\mathbf{q}\}$ we
get
\begin{eqnarray}
({\setbox0=\hbox{\thorn}\kern-.025em\copy0\kern-\wd0\kern.05em\copy0\kern-\wd%
0\kern-0.025em\raise.0433em\box0^{\prime }}{\setbox0=\hbox{$\eta$}\kern%
-.025em\copy0\kern-\wd0\kern.05em\copy0\kern-\wd0\kern-0.025em\raise.0433em%
\box0})\cdot {\setbox0=\hbox{o}\kern-.025em\copy0\kern-\wd0\kern.05em\copy0%
\kern-\wd0\kern-0.025em\raise.0433em\box0}={\scriptstyle{\frac{1}{3}}}\{{%
\setbox0=\hbox{\thorn}\kern-.025em\copy0\kern-\wd0\kern.05em\copy0\kern-\wd0%
\kern-0.025em\raise.0433em\box0^{\prime }}({\setbox0=\hbox{$\eta$}\kern%
-.025em\copy0\kern-\wd0\kern.05em\copy0\kern-\wd0\kern-0.025em\raise.0433em%
\box0}\cdot {\setbox0=\hbox{o}\kern-.025em\copy0\kern-\wd0\kern.05em\copy0%
\kern-\wd0\kern-0.025em\raise.0433em\box0})+({\setbox0=\hbox{\edth}\kern%
-.025em\copy0\kern-\wd0\kern.05em\copy0\kern-\wd0\kern-0.025em\raise.0433em%
\box0}{\setbox0=\hbox{$\eta$}\kern-.025em\copy0\kern-\wd0\kern.05em\copy0%
\kern-\wd0\kern-0.025em\raise.0433em\box0})-(\mathbf{p}-{1})\mathbf{T}{%
\setbox0=\hbox{$\eta$}\kern-.025em\copy0\kern-\wd0\kern.05em\copy0\kern-\wd0%
\kern-0.025em\raise.0433em\box0}\}
\label{bthornp.o}
\end{eqnarray}%
and
\begin{eqnarray}
({\setbox0=\hbox{\thorn}\kern-.025em\copy0\kern-\wd0\kern.05em\copy0\kern-\wd%
0\kern-0.025em\raise.0433em\box0^{\prime }}{\setbox0=\hbox{$\eta$}\kern%
-.025em\copy0\kern-\wd0\kern.05em\copy0\kern-\wd0\kern-0.025em\raise.0433em%
\box0})\cdot {\overline{\setbox0=\hbox{o}\kern-.025em\copy0\kern-\wd0\kern%
.05em\copy0\kern-\wd0\kern-0.025em\raise.0433em\box0}}={\scriptstyle{\frac{1%
}{3}}}\{{\setbox0=\hbox{\thorn}\kern-.025em\copy0\kern-\wd0\kern.05em\copy0%
\kern-\wd0\kern-0.025em\raise.0433em\box0^{\prime }}({\setbox0=\hbox{$\eta$}%
\kern-.025em\copy0\kern-\wd0\kern.05em\copy0\kern-\wd0\kern-0.025em\raise%
.0433em\box0}\cdot \overline{\setbox0=\hbox{o}\kern-.025em\copy0\kern-\wd0%
\kern.05em\copy0\kern-\wd0\kern-0.025em\raise.0433em\box0})+({\setbox0=%
\hbox{\edth}\kern-.025em\copy0\kern-\wd0\kern.05em\copy0\kern-\wd0\kern%
-0.025em\raise.0433em\box0}^{\prime }{\setbox0=\hbox{$\eta$}\kern-.025em\copy%
0\kern-\wd0\kern.05em\copy0\kern-\wd0\kern-0.025em\raise.0433em\box0})-%
\mathbf{q}\mathbf{\overline{T}}{\setbox0=\hbox{$\eta$}\kern-.025em\copy0\kern%
-\wd0\kern.05em\copy0\kern-\wd0\kern-0.025em\raise.0433em\box0}\}
\label{bthornp.ob}
\end{eqnarray}%

An alternative way to define the GIF operators is via the GHP operators $%
\hbox{\rm I}\kern-0.32em\raise0.35ex\hbox{\it o}, \hbox{$\partial$%
\kern-0.25em\raise0.6ex\hbox{\rm\char'40}},\hbox{$\partial$\kern-0.25em%
\raise0.6ex\hbox{\rm\char'40}}^{\prime}, \hbox{\rm I}\kern-0.32em\raise0.35ex%
\hbox{\it o}^{\prime}$, and in the case of a scalar field this gives
\begin{eqnarray}
({\setbox0=\hbox{\thorn}\kern-.025em\copy0\kern-\wd0\kern.05em\copy0\kern-\wd%
0\kern-0.025em\raise.0433em\box0^{\prime }}\eta )_{ABA^{\prime }B^{\prime }}
&=&(\hbox{\rm I}\kern-0.32em\raise0.35ex\hbox{\it o}^{\prime }\eta ){\setbox%
0=\hbox{o}\kern-.025em\copy0\kern-\wd0\kern.05em\copy0\kern-\wd0\kern-0.025em%
\raise.0433em\box0}_{A}{\setbox0=\hbox{o}\kern-.025em\copy0\kern-\wd0\kern%
.05em\copy0\kern-\wd0\kern-0.025em\raise.0433em\box0}_{B}{\overline{\setbox0=%
\hbox{o}\kern-.025em\copy0\kern-\wd0\kern.05em\copy0\kern-\wd0\kern-0.025em%
\raise.0433em\box0}}_{A^{\prime }}{\overline{\setbox0=\hbox{o}\kern-.025em%
\copy0\kern-\wd0\kern.05em\copy0\kern-\wd0\kern-0.025em\raise.0433em\box0}}%
_{B^{\prime }}-(\hbox{$\partial$\kern-0.25em\raise0.6ex\hbox{\rm\char'40}}%
^{\prime }\eta -q\bar{\tau}\eta ){\setbox0=\hbox{o}\kern-.025em\copy0\kern-%
\wd0\kern.05em\copy0\kern-\wd0\kern-0.025em\raise.0433em\box0}_{A}{\setbox0=%
\hbox{o}\kern-.025em\copy0\kern-\wd0\kern.05em\copy0\kern-\wd0\kern-0.025em%
\raise.0433em\box0}_{B}{\overline{\setbox0=\hbox{o}\kern-.025em\copy0\kern-%
\wd0\kern.05em\copy0\kern-\wd0\kern-0.025em\raise.0433em\box0}}_{(A^{\prime
}}{\overline{\setbox0=\hbox{$\iota$}\kern-.025em\copy0\kern-\wd0\kern.05em%
\copy0\kern-\wd0\kern-0.025em\raise.0433em\box0}}_{B^{\prime })}  \nonumber
\\
&&\ -(\hbox{$\partial$\kern-0.25em\raise0.6ex\hbox{\rm\char'40}}\eta -p\tau
\eta ){\setbox0=\hbox{o}\kern-.025em\copy0\kern-\wd0\kern.05em\copy0\kern-\wd%
0\kern-0.025em\raise.0433em\box0}_{(A}{\setbox0=\hbox{$\iota$}\kern-.025em%
\copy0\kern-\wd0\kern.05em\copy0\kern-\wd0\kern-0.025em\raise.0433em\box0}%
_{B)}{\overline{\setbox0=\hbox{o}\kern-.025em\copy0\kern-\wd0\kern.05em\copy0%
\kern-\wd0\kern-0.025em\raise.0433em\box0}}_{A^{\prime }}{\overline{\setbox0=%
\hbox{o}\kern-.025em\copy0\kern-\wd0\kern.05em\copy0\kern-\wd0\kern-0.025em%
\raise.0433em\box0}}_{B^{\prime }}+(\hbox{\rm I}\kern-0.32em\raise0.35ex%
\hbox{\it o}\eta -p\rho \eta -q\bar{\rho}\eta ){\setbox0=\hbox{o}\kern-.025em%
\copy0\kern-\wd0\kern.05em\copy0\kern-\wd0\kern-0.025em\raise.0433em\box0}%
_{(A}{\setbox0=\hbox{$\iota$}\kern-.025em\copy0\kern-\wd0\kern.05em\copy0%
\kern-\wd0\kern-0.025em\raise.0433em\box0}_{B)}{\overline{\setbox0=\hbox{o}%
\kern-.025em\copy0\kern-\wd0\kern.05em\copy0\kern-\wd0\kern-0.025em\raise%
.0433em\box0}}_{(A^{\prime }}{\overline{\setbox0=\hbox{$\iota$}\kern-.025em%
\copy0\kern-\wd0\kern.05em\copy0\kern-\wd0\kern-0.025em\raise.0433em\box0}}%
_{B^{\prime })}  \nonumber \\
&&\ -p\sigma {\setbox0=\hbox{$\iota$}\kern-.025em\copy0\kern-\wd0\kern.05em%
\copy0\kern-\wd0\kern-0.025em\raise.0433em\box0}_{A}{\setbox0=\hbox{$\iota$}%
\kern-.025em\copy0\kern-\wd0\kern.05em\copy0\kern-\wd0\kern-0.025em\raise%
.0433em\box0}_{B}{\overline{\setbox0=\hbox{o}\kern-.025em\copy0\kern-\wd0%
\kern.05em\copy0\kern-\wd0\kern-0.025em\raise.0433em\box0}}_{A^{\prime }}{%
\overline{\setbox0=\hbox{o}\kern-.025em\copy0\kern-\wd0\kern.05em\copy0\kern-%
\wd0\kern-0.025em\raise.0433em\box0}}_{B^{\prime }}-q\bar{\sigma}{\setbox0=%
\hbox{o}\kern-.025em\copy0\kern-\wd0\kern.05em\copy0\kern-\wd0\kern-0.025em%
\raise.0433em\box0}_{A}{\setbox0=\hbox{o}\kern-.025em\copy0\kern-\wd0\kern%
.05em\copy0\kern-\wd0\kern-0.025em\raise.0433em\box0}_{B}{\setbox0=%
\hbox{$\iota$}\kern-.025em\copy0\kern-\wd0\kern.05em\copy0\kern-\wd0\kern%
-0.025em\raise.0433em\box0}_{A^{\prime }}{\setbox0=\hbox{$\iota$}\kern-.025em%
\copy0\kern-\wd0\kern.05em\copy0\kern-\wd0\kern-0.025em\raise.0433em\box0}%
_{B^{\prime }}  \nonumber \\
&&\ +p\kappa {\setbox0=\hbox{$\iota$}\kern-.025em\copy0\kern-\wd0\kern.05em%
\copy0\kern-\wd0\kern-0.025em\raise.0433em\box0}_{A}{\setbox0=\hbox{$\iota$}%
\kern-.025em\copy0\kern-\wd0\kern.05em\copy0\kern-\wd0\kern-0.025em\raise%
.0433em\box0}_{B}{\overline{\setbox0=\hbox{o}\kern-.025em\copy0\kern-\wd0%
\kern.05em\copy0\kern-\wd0\kern-0.025em\raise.0433em\box0}}_{(A^{\prime }}{%
\overline{\setbox0=\hbox{$\iota$}\kern-.025em\copy0\kern-\wd0\kern.05em\copy0%
\kern-\wd0\kern-0.025em\raise.0433em\box0}}_{B^{\prime })}+q\bar{\kappa}{%
\setbox0=\hbox{o}\kern-.025em\copy0\kern-\wd0\kern.05em\copy0\kern-\wd0\kern%
-0.025em\raise.0433em\box0}_{(A}{\setbox0=\hbox{$\iota$}\kern-.025em\copy0%
\kern-\wd0\kern.05em\copy0\kern-\wd0\kern-0.025em\raise.0433em\box0}_{B)}{%
\overline{\setbox0=\hbox{$\iota$}\kern-.025em\copy0\kern-\wd0\kern.05em\copy0%
\kern-\wd0\kern-0.025em\raise.0433em\box0}}_{A^{\prime }}{\overline{\setbox0=%
\hbox{$\iota$}\kern-.025em\copy0\kern-\wd0\kern.05em\copy0\kern-\wd0\kern%
-0.025em\raise.0433em\box0}}_{B^{\prime }}  \label{thornp}
\end{eqnarray}%
\begin{eqnarray}
({\setbox0=\hbox{\edth}\kern-.025em\copy0\kern-\wd0\kern.05em\copy0\kern-\wd0%
\kern-0.025em\raise.0433em\box0^{\prime }}\eta )_{ABA^{\prime }} &=&(%
\hbox{$\partial$\kern-0.25em\raise0.6ex\hbox{\rm\char'40}}^{\prime }\eta ){%
\setbox0=\hbox{o}\kern-.025em\copy0\kern-\wd0\kern.05em\copy0\kern-\wd0\kern%
-0.025em\raise.0433em\box0}_{A}{\setbox0=\hbox{o}\kern-.025em\copy0\kern-\wd0%
\kern.05em\copy0\kern-\wd0\kern-0.025em\raise.0433em\box0}_{B}{\overline{%
\setbox0=\hbox{o}\kern-.025em\copy0\kern-\wd0\kern.05em\copy0\kern-\wd0\kern%
-0.025em\raise.0433em\box0}}_{A^{\prime }}-(\hbox{\rm I}\kern-0.32em\raise%
0.35ex\hbox{\it o}\eta -p\rho \eta ){\setbox0=\hbox{o}\kern-.025em\copy0\kern%
-\wd0\kern.05em\copy0\kern-\wd0\kern-0.025em\raise.0433em\box0}_{(A}{\setbox%
0=\hbox{$\iota$}\kern-.025em\copy0\kern-\wd0\kern.05em\copy0\kern-\wd0\kern%
-0.025em\raise.0433em\box0}_{B)}{\overline{\setbox0=\hbox{o}\kern-.025em\copy%
0\kern-\wd0\kern.05em\copy0\kern-\wd0\kern-0.025em\raise.0433em\box0}}%
_{A^{\prime }}  \nonumber \\
&&\ +q\bar{\sigma}{\setbox0=\hbox{o}\kern-.025em\copy0\kern-\wd0\kern.05em%
\copy0\kern-\wd0\kern-0.025em\raise.0433em\box0}_{A}{\setbox0=\hbox{o}\kern%
-.025em\copy0\kern-\wd0\kern.05em\copy0\kern-\wd0\kern-0.025em\raise.0433em%
\box0}_{B}{\setbox0=\hbox{$\iota$}\kern-.025em\copy0\kern-\wd0\kern.05em\copy%
0\kern-\wd0\kern-0.025em\raise.0433em\box0}_{A^{\prime }}-p\kappa {\setbox0=%
\hbox{$\iota$}\kern-.025em\copy0\kern-\wd0\kern.05em\copy0\kern-\wd0\kern%
-0.025em\raise.0433em\box0}_{A}{\setbox0=\hbox{$\iota$}\kern-.025em\copy0%
\kern-\wd0\kern.05em\copy0\kern-\wd0\kern-0.025em\raise.0433em\box0}_{B}{%
\overline{\setbox0=\hbox{o}\kern-.025em\copy0\kern-\wd0\kern.05em\copy0\kern-%
\wd0\kern-0.025em\raise.0433em\box0}}_{A^{\prime }}-q\bar{\kappa}{\setbox0=%
\hbox{o}\kern-.025em\copy0\kern-\wd0\kern.05em\copy0\kern-\wd0\kern-0.025em%
\raise.0433em\box0}_{(A}{\setbox0=\hbox{$\iota$}\kern-.025em\copy0\kern-\wd0%
\kern.05em\copy0\kern-\wd0\kern-0.025em\raise.0433em\box0}_{B)}{\overline{%
\setbox0=\hbox{$\iota$}\kern-.025em\copy0\kern-\wd0\kern.05em\copy0\kern-\wd0%
\kern-0.025em\raise.0433em\box0}}_{A^{\prime }}  \label{edthp}
\end{eqnarray}%
\begin{eqnarray}
({\setbox0=\hbox{\edth}\kern-.025em\copy0\kern-\wd0\kern.05em\copy0\kern-\wd0%
\kern-0.025em\raise.0433em\box0}\eta )_{AA^{\prime }B^{\prime }} &=&(%
\hbox{$\partial$\kern-0.25em\raise0.6ex\hbox{\rm\char'40}}\eta ){\setbox0=%
\hbox{o}\kern-.025em\copy0\kern-\wd0\kern.05em\copy0\kern-\wd0\kern-0.025em%
\raise.0433em\box0}_{A}{\overline{\setbox0=\hbox{o}\kern-.025em\copy0\kern-%
\wd0\kern.05em\copy0\kern-\wd0\kern-0.025em\raise.0433em\box0}}_{A^{\prime }}%
{\overline{\setbox0=\hbox{o}\kern-.025em\copy0\kern-\wd0\kern.05em\copy0\kern%
-\wd0\kern-0.025em\raise.0433em\box0}}_{B^{\prime }}-(\hbox{\rm I}\kern%
-0.32em\raise0.35ex\hbox{\it o}\eta -q\bar{\rho}\eta ){\setbox0=\hbox{o}\kern%
-.025em\copy0\kern-\wd0\kern.05em\copy0\kern-\wd0\kern-0.025em\raise.0433em%
\box0}_{A}{\overline{\setbox0=\hbox{o}\kern-.025em\copy0\kern-\wd0\kern.05em%
\copy0\kern-\wd0\kern-0.025em\raise.0433em\box0}}_{(A^{\prime }}{\overline{%
\setbox0=\hbox{$\iota$}\kern-.025em\copy0\kern-\wd0\kern.05em\copy0\kern-\wd0%
\kern-0.025em\raise.0433em\box0}}_{B^{\prime })}  \nonumber \\
&&\ +p\sigma {\setbox0=\hbox{$\iota$}\kern-.025em\copy0\kern-\wd0\kern.05em%
\copy0\kern-\wd0\kern-0.025em\raise.0433em\box0}_{A}{\overline{\setbox0=%
\hbox{o}\kern-.025em\copy0\kern-\wd0\kern.05em\copy0\kern-\wd0\kern-0.025em%
\raise.0433em\box0}}_{A^{\prime }}{\overline{\setbox0=\hbox{o}\kern-.025em%
\copy0\kern-\wd0\kern.05em\copy0\kern-\wd0\kern-0.025em\raise.0433em\box0}}%
_{B^{\prime }}-p\kappa {\setbox0=\hbox{$\iota$}\kern-.025em\copy0\kern-\wd0%
\kern.05em\copy0\kern-\wd0\kern-0.025em\raise.0433em\box0}_{A}{\overline{%
\setbox0=\hbox{o}\kern-.025em\copy0\kern-\wd0\kern.05em\copy0\kern-\wd0\kern%
-0.025em\raise.0433em\box0}}_{(A^{\prime }}{\overline{\setbox0=\hbox{$\iota$}%
\kern-.025em\copy0\kern-\wd0\kern.05em\copy0\kern-\wd0\kern-0.025em\raise%
.0433em\box0}}_{B^{\prime })}-q\bar{\kappa}{\setbox0=\hbox{o}\kern-.025em\copy%
0\kern-\wd0\kern.05em\copy0\kern-\wd0\kern-0.025em\raise.0433em\box0}_{A}{%
\overline{\setbox0=\hbox{$\iota$}\kern-.025em\copy0\kern-\wd0\kern.05em\copy0%
\kern-\wd0\kern-0.025em\raise.0433em\box0}}_{A^{\prime }}{\overline{\setbox0=%
\hbox{$\iota$}\kern-.025em\copy0\kern-\wd0\kern.05em\copy0\kern-\wd0\kern%
-0.025em\raise.0433em\box0}}_{B^{\prime }}  \label{edth}
\end{eqnarray}%
\begin{eqnarray}
({\setbox0=\hbox{\thorn}\kern-.025em\copy0\kern-\wd0\kern.05em\copy0\kern-\wd%
0\kern-0.025em\raise.0433em\box0}\eta )_{AA^{\prime }}=(\hbox{\rm I}\kern%
-0.32em\raise0.35ex\hbox{\it o}\eta ){\setbox0=\hbox{o}\kern-.025em\copy0%
\kern-\wd0\kern.05em\copy0\kern-\wd0\kern-0.025em\raise.0433em\box0}_{A}{%
\setbox0=\hbox{o}\kern-.025em\copy0\kern-\wd0\kern.05em\copy0\kern-\wd0\kern%
-0.025em\raise.0433em\box0}_{B}+p\kappa {\setbox0=\hbox{$\iota$}\kern-.025em%
\copy0\kern-\wd0\kern.05em\copy0\kern-\wd0\kern-0.025em\raise.0433em\box0}%
_{A}{\overline{\setbox0=\hbox{o}\kern-.025em\copy0\kern-\wd0\kern.05em\copy0%
\kern-\wd0\kern-0.025em\raise.0433em\box0}}_{A^{\prime }}-q\bar{\kappa}{%
\setbox0=\hbox{o}\kern-.025em\copy0\kern-\wd0\kern.05em\copy0\kern-\wd0\kern%
-0.025em\raise.0433em\box0}_{A}{\overline{\setbox0=\hbox{$\iota$}\kern-.025em%
\copy0\kern-\wd0\kern.05em\copy0\kern-\wd0\kern-0.025em\raise.0433em\box0}}%
_{A^{\prime }}\ .
\label{thorn}
\end{eqnarray}%
These identities will enable us to transfer from the GIF to GHP formalism in the next section.

\section{The equations}

\label{equations}

We are concerned with the Petrov type O pure radiation spaces with non-zero
Ricci scalar, and in fact we begin with identical equations to those  in \cite{edram2}, but we shall repeat them here for easy reference. In the usual way, we choose ${\setbox0=\hbox{o} \kern-.025em%
\copy0\kern-\wd0 \kern.05em\copy0\kern-\wd0 \kern-0.025em\raise.0433em\box0 }%
_A$ to be aligned with the propogation direction of the radiation, so that
the Ricci spinor takes the form
\begin{eqnarray}
{\setbox0=\hbox{$\Phi$} \kern-.025em\copy0\kern-\wd0 \kern.05em\copy0\kern-%
\wd0 \kern-0.025em\raise.0433em\box0 }_{ABA^{\prime}B^{\prime}}=\Phi{\setbox%
0=\hbox{o} \kern-.025em\copy0\kern-\wd0 \kern.05em\copy0\kern-\wd0 \kern%
-0.025em\raise.0433em\box0 }_A{\setbox0=\hbox{o} \kern-.025em\copy0\kern-\wd%
0 \kern.05em\copy0\kern-\wd0 \kern-0.025em\raise.0433em\box0 }_B{\overline{%
\setbox0=\hbox{o} \kern-.025em\copy0\kern-\wd0 \kern.05em\copy0\kern-\wd0 %
\kern-0.025em\raise.0433em\box0 }}_{A^{\prime}}{\overline{\setbox0=\hbox{o} %
\kern-.025em\copy0\kern-\wd0 \kern.05em\copy0\kern-\wd0 \kern-0.025em\raise%
.0433em\box0 }}_{B^{\prime}}  \label{Phi}
\end{eqnarray}
where $\Phi (=\Phi_{22})$ is a real scalar field of weight $\{2,2\}$; all
the other curvature components, except the Ricci scalar $\Lambda$, vanish.

For this class of spaces the well known property of the vanishing of the
spin coefficients $\kappa, \sigma, \rho$ means that in the GIF
\begin{eqnarray}
\mathbf{K} &=0  \nonumber  \label{K} \\
\mathbf{S} &=0  \nonumber  \label{S} \\
\mathbf{R} & =0  \label{R}
\end{eqnarray}
but
\begin{eqnarray}
\mathbf{T}_{AA^{\prime}}=\tau{\setbox0=\hbox{o} \kern-.025em\copy0\kern-\wd0 %
\kern.05em\copy0\kern-\wd0 \kern-0.025em\raise.0433em\box0 }_A{\overline{%
\setbox0=\hbox{o} \kern-.025em\copy0\kern-\wd0 \kern.05em\copy0\kern-\wd0 %
\kern-0.025em\raise.0433em\box0 }}_{A^{\prime}}
\end{eqnarray}
where the  scalar $\tau$ has weight $\{1,-1\}$.
Notice that  $\tau$  and $\Phi_{22}$  are both invariant under the group of
null rotations so that they can be used instead of their GIF spinor
equivalents; this gives a considerable simplification in the GIF notation.

The GIF equations are:

\medskip

(i) GIF Ricci equations:
\begin{eqnarray}  \label{ricci}
{\setbox0=\hbox{\thorn} \kern-.025em\copy0\kern-\wd0 \kern.05em\copy0\kern-%
\wd0 \kern-0.025em\raise.0433em\box0 } \tau &=& 0 \\
{\setbox0=\hbox{\edth} \kern-.025em\copy0\kern-\wd0 \kern.05em\copy0\kern-\wd%
0 \kern-0.025em\raise.0433em\box0 } \tau &=& \tau^2 \\
{{\setbox0=\hbox{\edth} \kern-.025em\copy0\kern-\wd0 \kern.05em\copy0\kern-%
\wd0 \kern-0.025em\raise.0433em\box0 }^\prime} \tau&=& \tau\overline{\tau}%
+2\Lambda
\end{eqnarray}

(ii) GIF Bianchi equations:
\begin{eqnarray}  \label{bianchi1}
{\setbox0=\hbox{\thorn} \kern-.025em\copy0\kern-\wd0 \kern.05em\copy0\kern-%
\wd0 \kern-0.025em\raise.0433em\box0 }\Phi &=& 0 \\
{\setbox0=\hbox{\edth} \kern-.025em\copy0\kern-\wd0 \kern.05em\copy0\kern-\wd%
0 \kern-0.025em\raise.0433em\box0 }\Phi &=& \tau\Phi \\
{{\setbox0=\hbox{\edth} \kern-.025em\copy0\kern-\wd0 \kern.05em\copy0\kern-%
\wd0 \kern-0.025em\raise.0433em\box0 }^\prime}\Phi &=& \overline{\tau}\Phi
\end{eqnarray}
\begin{eqnarray}  \label{bianchi2}
{\setbox0=\hbox{\thorn} \kern-.025em\copy0\kern-\wd0 \kern.05em\copy0\kern-%
\wd0 \kern-0.025em\raise.0433em\box0 }\Lambda &=& 0  \nonumber \\
{\setbox0=\hbox{\edth} \kern-.025em\copy0\kern-\wd0 \kern.05em\copy0\kern-\wd%
0 \kern-0.025em\raise.0433em\box0 }\Lambda &=& 0  \nonumber \\
{{\setbox0=\hbox{\edth} \kern-.025em\copy0\kern-\wd0 \kern.05em\copy0\kern-%
\wd0 \kern-0.025em\raise.0433em\box0 }^\prime}\Lambda &=& 0  \nonumber \\
{{\setbox0=\hbox{\thorn} \kern-.025em\copy0\kern-\wd0 \kern.05em\copy0\kern-%
\wd0 \kern-0.025em\raise.0433em\box0 }^\prime}\Lambda &=& 0
\end{eqnarray}

(iii) GIF commutators (applied to a general symmetric spinor {\boldmath$\eta$%
} of weight $\mathbf{\{p,q\}}$ and with $N$ unprimed and $N^{\prime}$ primed
indices):
\begin{eqnarray}  \label{commf}
({\setbox0=\hbox{\thorn} \kern-.025em\copy0\kern-\wd0 \kern.05em\copy0\kern-%
\wd0 \kern-0.025em\raise.0433em\box0 } {\setbox0=\hbox{\thorn} \kern-.025em%
\copy0\kern-\wd0 \kern.05em\copy0\kern-\wd0 \kern-0.025em\raise.0433em\box0 }%
^\prime - {\setbox0=\hbox{\thorn} \kern-.025em\copy0\kern-\wd0 \kern.05em%
\copy0\kern-\wd0 \kern-0.025em\raise.0433em\box0 }^\prime{\setbox0=%
\hbox{\thorn} \kern-.025em\copy0\kern-\wd0 \kern.05em\copy0\kern-\wd0 \kern%
-0.025em\raise.0433em\box0 })\mbox{\boldmath$\eta$} &=& (\overline{\tau}{%
\setbox0=\hbox{\edth} \kern-.025em\copy0\kern-\wd0 \kern.05em\copy0\kern-\wd%
0 \kern-0.025em\raise.0433em\box0 } + \tau{\setbox0=\hbox{\edth} \kern-.025em%
\copy0\kern-\wd0 \kern.05em\copy0\kern-\wd0 \kern-0.025em\raise.0433em\box0 }%
^\prime)\mbox{\boldmath$\eta$}+(\mathbf{p}-N)\Lambda \mbox{\boldmath$\eta$}+(%
\mathbf{q}-N^\prime)\Lambda \mbox{\boldmath$\eta$} \\
({\setbox0=\hbox{\thorn} \kern-.025em\copy0\kern-\wd0 \kern.05em\copy0\kern-%
\wd0 \kern-0.025em\raise.0433em\box0 }{\setbox0=\hbox{\edth} \kern-.025em%
\copy0\kern-\wd0 \kern.05em\copy0\kern-\wd0 \kern-0.025em\raise.0433em\box0 }
- {\setbox0=\hbox{\edth} \kern-.025em\copy0\kern-\wd0 \kern.05em\copy0\kern-%
\wd0 \kern-0.025em\raise.0433em\box0 }{\setbox0=\hbox{\thorn} \kern-.025em%
\copy0\kern-\wd0 \kern.05em\copy0\kern-\wd0 \kern-0.025em\raise.0433em\box0 }%
)\mbox{\boldmath$\eta$}&=& 2\Lambda(\mbox{\boldmath$\eta$}\cdot{\setbox0=%
\hbox{o} \kern-.025em\copy0\kern-\wd0 \kern.05em\copy0\kern-\wd0 \kern%
-0.025em\raise.0433em\box0 }) \\
({\setbox0=\hbox{\thorn} \kern-.025em\copy0\kern-\wd0 \kern.05em\copy0\kern-%
\wd0 \kern-0.025em\raise.0433em\box0 }{\setbox0=\hbox{\edth} \kern-.025em%
\copy0\kern-\wd0 \kern.05em\copy0\kern-\wd0 \kern-0.025em\raise.0433em\box0 }%
^{\prime} - {\setbox0=\hbox{\edth} \kern-.025em\copy0\kern-\wd0 \kern.05em%
\copy0\kern-\wd0 \kern-0.025em\raise.0433em\box0 }^{\prime}{\setbox0=%
\hbox{\thorn} \kern-.025em\copy0\kern-\wd0 \kern.05em\copy0\kern-\wd0 \kern%
-0.025em\raise.0433em\box0 })\mbox{\boldmath$\eta$}&=& 2\Lambda(%
\mbox{\boldmath$\eta$}\cdot \overline{{\setbox0=\hbox{o} \kern-.025em\copy0%
\kern-\wd0 \kern.05em\copy0\kern-\wd0 \kern-0.025em\raise.0433em\box0 }}) \\
({\setbox0=\hbox{\edth} \kern-.025em\copy0\kern-\wd0 \kern.05em\copy0\kern-%
\wd0 \kern-0.025em\raise.0433em\box0 }{\setbox0=\hbox{\edth} \kern-.025em%
\copy0\kern-\wd0 \kern.05em\copy0\kern-\wd0 \kern-0.025em\raise.0433em\box0 }%
^\prime - {\setbox0=\hbox{\edth} \kern-.025em\copy0\kern-\wd0 \kern.05em\copy%
0\kern-\wd0 \kern-0.025em\raise.0433em\box0 }^\prime{\setbox0=\hbox{\edth} %
\kern-.025em\copy0\kern-\wd0 \kern.05em\copy0\kern-\wd0 \kern-0.025em\raise%
.0433em\box0 })\mbox{\boldmath$\eta$} &=& -(\mathbf{p}-N)\Lambda %
\mbox{\boldmath$\eta$}+(\mathbf{q}-N^\prime)\Lambda \mbox{\boldmath$\eta$} \\
({\setbox0=\hbox{\thorn} \kern-.025em\copy0\kern-\wd0 \kern.05em\copy0\kern-%
\wd0 \kern-0.025em\raise.0433em\box0 }^\prime{\setbox0=\hbox{\edth} \kern%
-.025em\copy0\kern-\wd0 \kern.05em\copy0\kern-\wd0 \kern-0.025em\raise.0433em%
\box0 } - {\setbox0=\hbox{\edth} \kern-.025em\copy0\kern-\wd0 \kern.05em\copy%
0\kern-\wd0 \kern-0.025em\raise.0433em\box0 }{\setbox0=\hbox{\thorn} \kern%
-.025em\copy0\kern-\wd0 \kern.05em\copy0\kern-\wd0 \kern-0.025em\raise.0433em%
\box0 }^\prime)\mbox{\boldmath$\eta$} &=& -\tau{\setbox0=\hbox{\thorn} \kern%
-.025em\copy0\kern-\wd0 \kern.05em\copy0\kern-\wd0 \kern-0.025em\raise.0433em%
\box0 }^\prime\mbox{\boldmath$\eta$} -\Phi(\mbox{\boldmath$\eta$}\cdot {%
\setbox0=\hbox{o} \kern-.025em\copy0\kern-\wd0 \kern.05em\copy0\kern-\wd0 %
\kern-0.025em\raise.0433em\box0 }) \\
({\setbox0=\hbox{\thorn} \kern-.025em\copy0\kern-\wd0 \kern.05em\copy0\kern-%
\wd0 \kern-0.025em\raise.0433em\box0 }^\prime{\setbox0=\hbox{\edth} \kern%
-.025em\copy0\kern-\wd0 \kern.05em\copy0\kern-\wd0 \kern-0.025em\raise.0433em%
\box0 }^{\prime} - {\setbox0=\hbox{\edth} \kern-.025em\copy0\kern-\wd0 \kern%
.05em\copy0\kern-\wd0 \kern-0.025em\raise.0433em\box0 }^{\prime}{\setbox0=%
\hbox{\thorn} \kern-.025em\copy0\kern-\wd0 \kern.05em\copy0\kern-\wd0 \kern%
-0.025em\raise.0433em\box0 }^\prime)\mbox{\boldmath$\eta$} &=& -\overline{%
\tau}{\setbox0=\hbox{\thorn} \kern-.025em\copy0\kern-\wd0 \kern.05em\copy0%
\kern-\wd0 \kern-0.025em\raise.0433em\box0 }^\prime\mbox{\boldmath$\eta$}
-\Phi(\mbox{\boldmath$\eta$}\cdot \overline{{\setbox0=\hbox{o} \kern-.025em%
\copy0\kern-\wd0 \kern.05em\copy0\kern-\wd0 \kern-0.025em\raise.0433em\box0 }
})
\end{eqnarray}
where $(\mbox{\boldmath$\eta$} \cdot \mathbf{o})$ is the $(N-1,N^{\prime})$%
-spinor $\mbox{\boldmath$\eta$}_{A_1....A_NA_1....A_{N^{\prime}}}\mathbf{o}%
^{A_N}$ , and $(\mbox{\boldmath$\eta$} \cdot \mathbf{\bar o})$ is the $%
(N,N^{\prime}-1)$-spinor $\mbox{\boldmath$\eta$}_{A_1....A_NA_1....A_{N^{%
\prime}}}\bar {\mathbf{o}}^{A_{N^{\prime}}}$, and if the contraction is not
possible then these terms are set to zero.

These GIF equations contain all the information for the type O pure
radiation metrics with non-zero Ricci scalar. We emphasize that we assume
throughout that constant $\Lambda\neq 0$ as well as $\tau\neq 0$.

We noted in \cite{edram2} that the type O pure
radiation metrics with non-zero Ricci scalar divided naturally into two cases

(i) $\Lambda +\tau\bar\tau \ne 0$

(ii) $\Lambda +\tau\bar\tau = 0$

In \cite{edram2} we considered the second of these subclasses; in the present paper we consider the first.

It will be convenient to introduce
\begin{equation}
{k\!\!\!\!^{-} {}}\equiv (\Lambda +\tau\overline\tau)/2\tau\overline\tau
\label{A1defn}
\end{equation}
for notational convenience\footnote{%
This quantity ${k\!\!\!\!^{-} {}}$ is closely related to the quantity $%
\kappa $ in \cite{ozs} and to $k$ in \cite{gr1}; any of these quantities can be used to classify the conformally flat pure radiation spaces (as well as more general Petrov types) into different subclasses.}, and so throughout this
paper we will assume
\[
{k\!\!\!\!^{-} {}}\ne 0 \, .
\]

\section{Transfering from GIF to the GHP formalism}

\label{transfer}

As emphasised in the Introduction, calculations in the GIF can be long and
complicated, and a careful examination of the details of \cite{edvic}
reveals that there is some redundancy in the techniques introduced there. In
particular, the complete and involutive set of tables for all of the scalar
quantities were obtained in terms of GIF operators, whereas we really only
need the simpler GHP version of the tables in order to deduce the metric.
In fact, in a particular calculation such as \cite{edvic} and in the present paper, once we have used
the GIF to obtain the table for the second spinor $\mathbf{I}$, and applied
the GIF commutators to this table, we can then identify $\mathbf{I}$ with
the second dyad spinor ${\setbox0=\hbox{$\iota$} \kern-.025em\copy0\kern-\wd%
0 \kern.05em\copy0\kern-\wd0 \kern-0.025em\raise.0433em\box0 }$ in the GHP
formalism and \textit{immediately} translate any existing tables into the
GHP formalism, which means that we can  then complete the calculations for the remaining tables
in the GHP formalism.

For the subsequent calculations we will need the GHP commutator equations,
which can be obtained from the GIF commutator equations by projecting on the
appropriate number of ${\setbox0=\hbox{$\iota$}\kern-.025em\copy0\kern-\wd0%
\kern.05em\copy0\kern-\wd0\kern-0.025em\raise.0433em\box0}^{A},\bar {\setbox0=\hbox{$\iota$}\kern-.025em\copy0\kern-\wd0%
\kern.05em\copy0\kern-\wd0\kern-0.025em\raise.0433em\box0}
^{A^{\prime }}$ spinors and using (\ref{thornp}), (\ref{edthp}), (\ref{edth}%
), (\ref{thorn}). An alternative method, which will be quicker for our purpose, is to make use of the GHP commutator equations as quoted in
\cite{ghp} (specialised to this class of spacetimes),
\begin{eqnarray}
(\hbox{\rm I}\kern-0.32em\raise0.35ex\hbox{\it o}\hbox{\rm I}\kern-0.32em%
\raise0.35ex\hbox{\it o}^{\prime }-\hbox{\rm I}\kern-0.32em\raise0.35ex%
\hbox{\it o}^{\prime }\hbox{\rm I}\kern-0.32em\raise0.35ex\hbox{\it o})\eta
&=&\Bigl((\bar{\tau}-\tau ^{\prime })\hbox{$\partial$\kern-0.25em\raise0.6ex%
\hbox{\rm\char'40}}+(\tau -\bar{\tau}^{\prime })\hbox{$\partial$\kern-0.25em%
\raise0.6ex\hbox{\rm\char'40}}^{\prime }+p(\tau \tau ^{\prime }+\Lambda )+q(%
\bar{\tau}\bar{\tau}^{\prime }+\Lambda )\Bigr)\eta   \nonumber
\label{GHPcomm} \\
(\hbox{\rm I}\kern-0.32em\raise0.35ex\hbox{\it o}\hbox{$\partial$%
\kern-0.25em\raise0.6ex\hbox{\rm\char'40}}-\hbox{$\partial$\kern-0.25em%
\raise0.6ex\hbox{\rm\char'40}}\hbox{\rm I}\kern-0.32em\raise0.35ex%
\hbox{\it
o})\eta  &=&-\bar{\tau}^{\prime }\hbox{\rm I}\kern-0.32em\raise0.35ex%
\hbox{\it o}\eta   \nonumber \\
(\hbox{$\partial$\kern-0.25em\raise0.6ex\hbox{\rm\char'40}}%
\hbox{$\partial$\kern-0.25em\raise0.6ex\hbox{\rm\char'40}}^{\prime }-%
\hbox{$\partial$\kern-0.25em\raise0.6ex\hbox{\rm\char'40}}^{\prime }%
\hbox{$\partial$\kern-0.25em\raise0.6ex\hbox{\rm\char'40}})\eta  &=&\Bigl(%
(\bar{\rho}^{\prime }-{\rho}^{\prime }) \hbox{\rm I}\kern-0.32em\raise0.35ex\hbox{\it o}%
-p\Lambda +q\Lambda \Bigr)\eta   \nonumber \\
(\hbox{\rm I}\kern-0.32em\raise0.35ex\hbox{\it o}^{\prime }%
\hbox{$\partial$\kern-0.25em\raise0.6ex\hbox{\rm\char'40}}-%
\hbox{$\partial$\kern-0.25em\raise0.6ex\hbox{\rm\char'40}}\hbox{\rm I}\kern%
-0.32em\raise0.35ex\hbox{\it o}^{\prime })\eta  &=&\Bigl(\rho ^{\prime }%
\hbox{$\partial$\kern-0.25em\raise0.6ex\hbox{\rm\char'40}}+\bar{\sigma}%
^{\prime }\hbox{$\partial$\kern-0.25em\raise0.6ex\hbox{\rm\char'40}}^{\prime
}-\tau \hbox{\rm I}\kern-0.32em\raise0.35ex\hbox{\it o}^{\prime }-\bar{\kappa%
}^{\prime }\hbox{\rm I}\kern-0.32em\raise0.35ex\hbox{\it o}-q\bar{\tau}\bar{%
\sigma}^{\prime }-p\rho ^{\prime }\tau \Bigr)\eta
\end{eqnarray}%
where $\eta $ is an arbitrary scalar of weight $\{p,q\}$.

Of course now we encounter the problem that the GHP formalism involves the
spin coefficients $\tau ^{\prime },\sigma ^{\prime },\mu ^{\prime },\kappa
^{\prime }$ which are missing from the GIF. However, assuming that we have
obtained a table for $\mathbf{I}$ in our GIF analysis, once we have
identified $\mathbf{I}$ with the second dyad spinor ${\setbox0=\hbox{$\iota$}%
\kern-.025em\copy0\kern-\wd0\kern.05em\copy0\kern-\wd0\kern-0.025em\raise%
.0433em\box0}$, we can use this table to obtain directly these additional
four spin coefficients as follows
\begin{eqnarray}
\tau ^{\prime }&=&-{\setbox0=\hbox{$\iota$}\kern-.025em\copy0\kern-\wd0\kern%
.05em\copy0\kern-\wd0\kern-0.025em\raise.0433em\box0}^{B}D\,({\setbox0=%
\hbox{$\iota$}\kern-.025em\copy0\kern-\wd0\kern.05em\copy0\kern-\wd0\kern%
-0.025em\raise.0433em\box0}_{B})=-{\setbox0=\hbox{$\iota$}\kern-.025em\copy0%
\kern-\wd0\kern.05em\copy0\kern-\wd0\kern-0.025em\raise.0433em\box0}^{B}%
\hbox{\rm I}\kern-0.32em\raise0.35ex\hbox{\it o}\,({\setbox0=\hbox{$\iota$}%
\kern-.025em\copy0\kern-\wd0\kern.05em\copy0\kern-\wd0\kern-0.025em\raise%
.0433em\box0}_{B})=-{\setbox0=\hbox{$\iota$}\kern-.025em\copy0\kern-\wd0\kern%
.05em\copy0\kern-\wd0\kern-0.025em\raise.0433em\box0}^{B}{\setbox0=%
\hbox{$\iota$}\kern-.025em\copy0\kern-\wd0\kern.05em\copy0\kern-\wd0\kern%
-0.025em\raise.0433em\box0}^{C}\bar{\imath}^{C^{\prime }}{\setbox0=%
\hbox{\thorn}\kern-.025em\copy0\kern-\wd0\kern.05em\copy0\kern-\wd0\kern%
-0.025em\raise.0433em\box0}_{CC^{\prime }}\,({\setbox0=\hbox{$\iota$}\kern%
-.025em\copy0\kern-\wd0\kern.05em\copy0\kern-\wd0\kern-0.025em\raise.0433em%
\box0}_{B})
 \nonumber \\
\rho ^{\prime } &=&-{\setbox0=\hbox{$\iota$}\kern-.025em\copy0\kern-\wd0\kern%
.05em\copy0\kern-\wd0\kern-0.025em\raise.0433em\box0}^{B}{\setbox0=%
\hbox{$\iota$}\kern-.025em\copy0\kern-\wd0\kern.05em\copy0\kern-\wd0\kern%
-0.025em\raise.0433em\box0}^{C}\bar{\imath}^{C^{\prime }}\bar{\imath}%
^{D^{\prime }}{\setbox0=\hbox{\edth}\kern-.025em\copy0\kern-\wd0\kern.05em%
\copy0\kern-\wd0\kern-0.025em\raise.0433em\box0}_{CC^{\prime }D^{\prime }}\,(%
{\setbox0=\hbox{$\iota$}\kern-.025em\copy0\kern-\wd0\kern.05em\copy0\kern-\wd%
0\kern-0.025em\raise.0433em\box0}_{B})\   \nonumber \\
\sigma ^{\prime } &=&-{\setbox0=\hbox{$\iota$}\kern-.025em\copy0\kern-\wd0%
\kern.05em\copy0\kern-\wd0\kern-0.025em\raise.0433em\box0}^{B}{\setbox0=%
\hbox{$\iota$}\kern-.025em\copy0\kern-\wd0\kern.05em\copy0\kern-\wd0\kern%
-0.025em\raise.0433em\box0}^{C}{\setbox0=\hbox{$\iota$}\kern-.025em\copy0%
\kern-\wd0\kern.05em\copy0\kern-\wd0\kern-0.025em\raise.0433em\box0}^{D}\bar{%
\imath}^{C^{\prime }}{\setbox0=\hbox{\edth}\kern-.025em\copy0\kern-\wd0\kern%
.05em\copy0\kern-\wd0\kern-0.025em\raise.0433em\box0}^{\prime }_{CDC^{\prime }}\,({%
\setbox0=\hbox{$\iota$}\kern-.025em\copy0\kern-\wd0\kern.05em\copy0\kern-\wd0%
\kern-0.025em\raise.0433em\box0}_{B})  \nonumber \\
\kappa ^{\prime } &=&-{\setbox0=\hbox{$\iota$}\kern-.025em\copy0\kern-\wd0%
\kern.05em\copy0\kern-\wd0\kern-0.025em\raise.0433em\box0}^{B}{\setbox0=%
\hbox{$\iota$}\kern-.025em\copy0\kern-\wd0\kern.05em\copy0\kern-\wd0\kern%
-0.025em\raise.0433em\box0}^{C}{\setbox0=\hbox{$\iota$}\kern-.025em\copy0%
\kern-\wd0\kern.05em\copy0\kern-\wd0\kern-0.025em\raise.0433em\box0}^{D}\bar{%
\imath}^{C^{\prime }}\bar{\imath}^{D^{\prime }}{{\setbox0=\hbox{\thorn}\kern%
-.025em\copy0\kern-\wd0\kern.05em\copy0\kern-\wd0\kern-0.025em\raise.0433em%
\box0}^{\prime }}_{CDC^{\prime }D^{\prime }}\ ({\setbox0=\hbox{$\iota$}\kern%
-.025em\copy0\kern-\wd0\kern.05em\copy0\kern-\wd0\kern-0.025em\raise.0433em%
\box0}_{B})  \label{rho'}
\end{eqnarray}

\section{The integration procedure for $\Lambda + \protect\tau\overline
\protect\tau \ne 0$: the generic case.}

\label{intne}

\subsection{Preliminary rearrangement.}

\label{intne1}

The Riemann tensor and the spin coefficients supply three real scalars which
can easily be rearranged to give one real zero-weighted $(\tau \overline{%
\tau })$ and two real weighted scalars, $\Phi $ and $\arg (\tau /\overline{%
\tau })$. However, in order to keep the presentation of subsequent
calculations to a minimum and to have easy comparison with \cite{edvic}, it
will be convenient to rearrange slightly these three scalars, and use
instead the zero-weighted scalar
\begin{equation}
A=\frac{1}{\sqrt{2\tau \overline{\tau }}}  \label{A}
\end{equation}%
and the weighted scalars\footnote{%
We have retained the notation $P,Q$ which was used in \cite{edvic} for these two
weighted scalars;  note the slightly different definitions compared with  ${\cal P}, {\cal Q}$ used in \cite{edram2} when considering the case $\Lambda + \tau\bar \tau =0$.  Care needs to be taken when comparing with  the various quantities labelled with $P,Q$
(sometimes $p,q$) in  \cite{ozs}, \cite{gr1}, \cite{bic1},
\cite{bic2}, \cite{edram1} and other references.}
\begin{equation}
P=\sqrt{\frac{\tau }{2\overline{\tau }}},  \label{P}
\end{equation}%
\begin{equation}
{Q}=\frac{\sqrt{\Phi }}{\sqrt[4]{2\tau \overline{\tau }}}
\label{Q}
\end{equation}%
where $P$ is a complex scalar of weight $\{1,-1\}$, with $
P\overline{P}={\frac{1}{2}}$;  and ${Q}$ is a real
scalar of weight $\{-1,-1\}$. (As well as $\Phi ={\frac{{Q}^{2}}{%
A}}\neq 0\neq \Lambda $, we are assuming $\tau =P/A\neq 0$, and so
each of $A,\ P,\ {Q}$, will always be defined and different from
zero.)

These particular choices enable us to replace the Ricci equations with
\begin{eqnarray}
{\setbox0=\hbox{\thorn}\kern-.025em\copy0\kern-\wd0\kern.05em\copy0\kern-\wd0%
\kern-0.025em\raise.0433em\box0}A &=&0  \nonumber  \label{tableA} \\
{\setbox0=\hbox{\edth}\kern-.025em\copy0\kern-\wd0\kern.05em\copy0\kern-\wd0%
\kern-0.025em\raise.0433em\box0}A &=&-2P(\Lambda A^{2}+1/2)=-2%
P{k\!\!\!\!^{-}{}}  \nonumber \\
{{\setbox0=\hbox{\edth}\kern-.025em\copy0\kern-\wd0\kern.05em\copy0\kern-\wd0%
\kern-0.025em\raise.0433em\box0}^{\prime }}A &=&-2\overline{P}%
(\Lambda A^{2}+1/2)=-2\overline{P}{k\!\!\!\!^{-}{}}
\label{partialtableA}
\end{eqnarray}%
\begin{eqnarray}
{\setbox0=\hbox{\thorn}\kern-.025em\copy0\kern-\wd0\kern.05em\copy0\kern-\wd0%
\kern-0.025em\raise.0433em\box0}(\overline{P}{Q}) &=&0
\nonumber  \label{partialtablePbarQ} \\
{\setbox0=\hbox{\edth}\kern-.025em\copy0\kern-\wd0\kern.05em\copy0\kern-\wd0%
\kern-0.025em\raise.0433em\box0}(\overline{P}{Q}) &=&%
\frac{1}{2}{Q}\Lambda A  \nonumber \\
{{\setbox0=\hbox{\edth}\kern-.025em\copy0\kern-\wd0\kern.05em\copy0\kern-\wd0%
\kern-0.025em\raise.0433em\box0}^{\prime }}(\overline{P}{\mathcal{Q%
}}) &=&-3{Q}\overline{P}^{2}\Lambda A
\label{partialtablePQ}
\end{eqnarray}%
where we now have
\[
{k\!\!\!\!^{-}{}}=\Lambda A^{2}+1/2\neq 0
\]%
At various steps in the sequel it will be obvious that we are assuming ${%
k\!\!\!\!^{-}{}}\neq 3/2$; however, this is not an additional restriction
since we can deduce from the partial table (\ref{partialtableA}) for $A$
that this condition must always be satisfied.

\subsection{Constructing a table for $\mathbf{I}$ and applying commutators
to $\mathbf{I}$.}

\label{intne2}

For our integration procedure we begin by completing the partial table (\ref%
{partialtablePQ}) for the $\{-2,0\}$ weighted scalar $\overline{P}%
Q$,
\begin{eqnarray}  \label{tablePbarQ}
{\setbox0=\hbox{\thorn} \kern-.025em\copy0\kern-\wd0 \kern.05em\copy0\kern-%
\wd0 \kern-0.025em\raise.0433em\box0 } (\overline {P}{Q}%
) &=& 0  \nonumber \\
{\setbox0=\hbox{\edth} \kern-.025em\copy0\kern-\wd0 \kern.05em\copy0\kern-\wd%
0 \kern-0.025em\raise.0433em\box0 } (\overline {P}{Q})&=&%
\frac{1}{2}\Lambda A{Q}  \nonumber \\
{{\setbox0=\hbox{\edth} \kern-.025em\copy0\kern-\wd0 \kern.05em\copy0\kern-%
\wd0 \kern-0.025em\raise.0433em\box0 }^\prime} (\overline {P}{%
Q}) &=&-3\Lambda A{Q}\overline{P}^2  \nonumber
\\
{{\setbox0=\hbox{\thorn} \kern-.025em\copy0\kern-\wd0 \kern.05em\copy0\kern-%
\wd0 \kern-0.025em\raise.0433em\box0 }^\prime} (\overline {P}{%
Q}) &=& \mathbf{J}  \label{tablePQ}
\end{eqnarray}
where we have completed the table with some spinor $\mathbf{J}$, which is as
yet undetermined.

We know from (\ref{thornp.ob}) and (\ref{thornp.o}) that
\begin{equation}
{{\setbox0=\hbox{\thorn}\kern-.025em\copy0\kern-\wd0\kern.05em\copy0\kern-\wd%
0\kern-0.025em\raise.0433em\box0}^{\prime }}(\overline{P}{\mathcal{%
Q}})\cdot {\overline{\mathbf{o}}}={{\setbox0=\hbox{\edth}\kern-.025em\copy0%
\kern-\wd0\kern.05em\copy0\kern-\wd0\kern-0.025em\raise.0433em\box0}^{\prime
}}(\overline{P}{Q})  \label{cdotobar}
\end{equation}%
\begin{equation}
{{\setbox0=\hbox{\thorn}\kern-.025em\copy0\kern-\wd0\kern.05em\copy0\kern-\wd%
0\kern-0.025em\raise.0433em\box0}^{\prime }}(\overline{P}{\mathcal{%
Q}})\cdot \mathbf{o}={\setbox0=\hbox{\edth}\kern-.025em\copy0\kern-\wd0\kern%
.05em\copy0\kern-\wd0\kern-0.025em\raise.0433em\box0}(\overline{P}{%
Q})+2\tau \overline{P}{Q}={\setbox0=%
\hbox{\edth}\kern-.025em\copy0\kern-\wd0\kern.05em\copy0\kern-\wd0\kern%
-0.025em\raise.0433em\box0}(\overline{P}{Q})+\frac{Q}{A}\label{cdoto}
\end{equation}%
Substituting (\ref{tablePQ}) we can then write
\begin{equation}
\mathbf{J}=-\Bigl(\frac{{Q}}{A}+\frac{1}{2}\Lambda A{Q}%
\Bigr)\mathbf{I}+3\Lambda A{Q}\overline{P}^{2}\overline{%
\mathbf{I}}\label{Jdefn}
\end{equation}%
where
\begin{equation}
\mathbf{I}\cdot {\overline{\setbox0=\hbox{o}\kern-.025em\copy0\kern-\wd0\kern%
.05em\copy0\kern-\wd0\kern-0.025em\raise.0433em\box0}}=0  \label{Iobar}
\end{equation}%
and
\begin{equation}
\mathbf{I}\cdot {\setbox0=\hbox{o}\kern-.025em\copy0\kern-\wd0\kern.05em\copy%
0\kern-\wd0\kern-0.025em\raise.0433em\box0}=-1  \label{Io}
\end{equation}%
Hence $\mathbf{I}$ is a $(1,0)$ valence spinor, and from
\begin{equation}
\Bigl({\setbox0=\hbox{\thorn}\kern-.025em\copy0\kern-\wd0\kern.05em\copy0%
\kern-\wd0\kern-0.025em\raise.0433em\box0}^{\prime }(\overline{P}{%
Q})\Bigr)_{ABA^{\prime }B^{\prime }}=-\Bigl({\frac{{Q}}{A%
}}+\frac{1}{2}\Lambda A{Q}\Bigr)\mathbf{I}_{(A}{\setbox0=\hbox{o}%
\kern-.025em\copy0\kern-\wd0\kern.05em\copy0\kern-\wd0\kern-0.025em\raise%
.0433em\box0}_{B)}{\overline{\setbox0=\hbox{o}\kern-.025em\copy0\kern-\wd0%
\kern.05em\copy0\kern-\wd0\kern-0.025em\raise.0433em\box0}}_{A^{\prime }}{%
\overline{\setbox0=\hbox{o}\kern-.025em\copy0\kern-\wd0\kern.05em\copy0\kern-%
\wd0\kern-0.025em\raise.0433em\box0}}_{B^{\prime }}+3\Lambda A{Q}%
\overline{P}^{2}\overline{\mathbf{I}}_{(A^{\prime }}{\overline{%
\setbox0=\hbox{o}\kern-.025em\copy0\kern-\wd0\kern.05em\copy0\kern-\wd0\kern%
-0.025em\raise.0433em\box0}}_{B^{\prime })}{\setbox0=\hbox{o}\kern-.025em%
\copy0\kern-\wd0\kern.05em\copy0\kern-\wd0\kern-0.025em\raise.0433em\box0}%
_{A}{\setbox0=\hbox{o}\kern-.025em\copy0\kern-\wd0\kern.05em\copy0\kern-\wd0%
\kern-0.025em\raise.0433em\box0}_{B}\label{Idefn}
\end{equation}%
we conclude that its weight is $\{-\setbox0=\hbox{1}\kern-.025em\copy0\kern-%
\wd0\kern.05em\copy0\kern-\wd0\kern-0.025em\raise.0433em\box0,\setbox0=%
\hbox{0}\kern-.025em\copy0\kern-\wd0\kern.05em\copy0\kern-\wd0\kern-0.025em%
\raise.0433em\box0\}$.

It is important to note two crucial properties of the new spinor $\mathbf{I}$. Firstly, for this whole class of spaces, $\mathbf{I}$ can never be zero,
nor parallel to ${\setbox0=\hbox{o}\kern-.025em\copy0\kern-\wd0\kern.05em%
\copy0\kern-\wd0\kern-0.025em\raise.0433em\box0}$. Secondly, for the
subclass under consideration in this paper,
the spinor $\mathbf{I}$ is given \textit{uniquely} in terms of the elements of the
GIF formalism and so is an intrinsic spinor; this can be seen when we solve for $\mathbf{I}$ from (\ref{Idefn}) and its complex conjugate remembering that  ${k\!\!\!\!^{-}{}}\neq 0$ in this paper.

(However, it is emphasised, on the contrary, for the subclass of spaces defined by ${%
k\!\!\!\!^{-}{}}=0$, that the spinor $\mathbf{I}$ is \textit{not} given
uniquely in terms of the elements of the GIF formalism and so is not an
intrinsic spinor; this is the subclass of spaces considered in  \cite{edram2}%
.)

\smallskip

It will be useful in the sequel to have separate tables for $P$
and ${Q}$ which are easily determined from (\ref{tablePQ}) as
follows:
\begin{eqnarray}  \label{completetableP}
{\setbox0=\hbox{\thorn} \kern-.025em\copy0\kern-\wd0 \kern.05em\copy0\kern-%
\wd0 \kern-0.025em\raise.0433em\box0 } P &=&0  \nonumber \\
{\setbox0=\hbox{\edth} \kern-.025em\copy0\kern-\wd0 \kern.05em\copy0\kern-\wd%
0 \kern-0.025em\raise.0433em\box0 } P &=& -2\Lambda AP^2
\nonumber \\
{{\setbox0=\hbox{\edth} \kern-.025em\copy0\kern-\wd0 \kern.05em\copy0\kern-%
\wd0 \kern-0.025em\raise.0433em\box0 }^\prime} P &=& \Lambda A
\nonumber \\
{{\setbox0=\hbox{\thorn} \kern-.025em\copy0\kern-\wd0 \kern.05em\copy0\kern-%
\wd0 \kern-0.025em\raise.0433em\box0 }^\prime} P &=& \frac{%
2P^2 {k\!\!\!\!^{-} {}}}{A}\mathbf{I} -\frac{{k\!\!\!\!^{-} {}}}{A}%
\overline{\mathbf{I}}
\end{eqnarray}
\begin{eqnarray}  \label{tableQ}
{\setbox0=\hbox{\thorn} \kern-.025em\copy0\kern-\wd0 \kern.05em\copy0\kern-%
\wd0 \kern-0.025em\raise.0433em\box0 } {Q} &=& 0  \nonumber \\
{\setbox0=\hbox{\edth} \kern-.025em\copy0\kern-\wd0 \kern.05em\copy0\kern-\wd%
0 \kern-0.025em\raise.0433em\box0 } {Q}&=&-\Lambda A{Q}%
P  \nonumber \\
{{\setbox0=\hbox{\edth} \kern-.025em\copy0\kern-\wd0 \kern.05em\copy0\kern-%
\wd0 \kern-0.025em\raise.0433em\box0 }^\prime} {Q} &=&-\Lambda A{%
Q}\overline{P}  \nonumber \\
{{\setbox0=\hbox{\thorn} \kern-.025em\copy0\kern-\wd0 \kern.05em\copy0\kern-%
\wd0 \kern-0.025em\raise.0433em\box0 }^\prime} {Q} &=& -\frac{{%
Q}P(\frac{3}{2}-{k\!\!\!\!^{-} {}})}{A}\mathbf{I} -\frac{%
{Q}\overline{P}(\frac{3}{2}-{k\!\!\!\!^{-} {}})}{A}%
\overline{\mathbf{I}}
\end{eqnarray}

Our first mission is to find the table for $\mathbf{I}$ which should follow
from applying the commutators to the table for $(\overline{P}{%
Q})$; but first of all we will need to complete the partial table (%
\ref{partialtableA}) for $A$ as an expression for ${{\setbox0=\hbox{\thorn}%
\kern-.025em\copy0\kern-\wd0\kern.05em\copy0\kern-\wd0\kern-0.025em\raise%
.0433em\box0}^{\prime }}A$ will be required. We obtain
\begin{eqnarray}
{\setbox0=\hbox{\thorn}\kern-.025em\copy0\kern-\wd0\kern.05em\copy0\kern-\wd0%
\kern-0.025em\raise.0433em\box0}A &=&0  \nonumber  \label{completetableA} \\
{\setbox0=\hbox{\edth}\kern-.025em\copy0\kern-\wd0\kern.05em\copy0\kern-\wd0%
\kern-0.025em\raise.0433em\box0}A &=&-2P{k\!\!\!\!^{-}{}}
\nonumber \\
{{\setbox0=\hbox{\edth}\kern-.025em\copy0\kern-\wd0\kern.05em\copy0\kern-\wd0%
\kern-0.025em\raise.0433em\box0}^{\prime }}A &=&-2\overline{P}{%
k\!\!\!\!^{-}{}}  \nonumber \\
{{\setbox0=\hbox{\thorn}\kern-.025em\copy0\kern-\wd0\kern.05em\copy0\kern-\wd%
0\kern-0.025em\raise.0433em\box0}^{\prime }}A &=&
\mathbf{C}
\end{eqnarray}%
where we have completed the table with a spinor $\mathbf{C}$, which is as
yet undetermined. It follows from (\ref{thornp.ob}) and (\ref{thornp.o}) that
\begin{equation}
\mathbf{C}\cdot \overline{{\setbox0=\hbox{o}\kern-.025em\copy0\kern-\wd0\kern%
.05em\copy0\kern-\wd0\kern-0.025em\raise.0433em\box0}}=
({{\setbox%
0=\hbox{\thorn}\kern-.025em\copy0\kern-\wd0\kern.05em\copy0\kern-\wd0\kern%
-0.025em\raise.0433em\box0}^{\prime }}A)\cdot \overline{{\setbox0=%
\hbox{o}\kern-.025em\copy0\kern-\wd0\kern.05em\copy0\kern-\wd0\kern-0.025em%
\raise.0433em\box0}}=(\bedth ^{\prime }A)=-2\overline{P}{%
k\!\!\!\!^{-}{}}  \label{A1}
\end{equation}%
\begin{equation}
\mathbf{C}\cdot {\setbox0=\hbox{o}\kern-.025em\copy0\kern-\wd0\kern.05em\copy%
0\kern-\wd0\kern-0.025em\raise.0433em\box0}=({{\setbox%
0=\hbox{\thorn}\kern-.025em\copy0\kern-\wd0\kern.05em\copy0\kern-\wd0\kern%
-0.025em\raise.0433em\box0}^{\prime }}A)\cdot {\setbox0=\hbox{o}\kern-.025em%
\copy0\kern-\wd0\kern.05em\copy0\kern-\wd0\kern-0.025em\raise.0433em\box0}=%
(\bedth A)=-2P{k\!\!\!\!^{-}{}}  \label{A2}
\end{equation}%
Therefore
\begin{equation}
\mathbf{C}=\frac{{Q}}{A}C{{k\!\!\!\!^{-}{}}}^{2}+2P{%
k\!\!\!\!^{-}{}}\mathbf{I}+2\overline{P}{%
k\!\!\!\!^{-}{}}\overline{\mathbf{I}}  \label{A3}
\end{equation}%
and so $\mathbf{C}$ is a Hermitian $(1,1)$ type spinor of weight $\{\mathbf{2}%
,\mathbf{2}\}$, with $C$  a zero-weighted real scalar, as yet undetermined (with the factor ${{%
k\!\!\!\!^{-}{}}}^{2}/A$ introduced for brevity and convenience of subsequent
presentation).

We are now able to apply the commutators to the table for $( \overline {%
P} {Q})$ which yields a partial table for the spinor $%
\mathbf{I}$; we obtain\begin{eqnarray}
{\setbox0=\hbox{\thorn} \kern-.025em\copy0\kern-\wd0 \kern.05em\copy0\kern-%
\wd0 \kern-0.025em\raise.0433em\box0 } \mathbf{I} &=& \frac{3\Lambda A%
\overline{P}}{(\frac{3}{2}-{k\!\!\!\!^{-} {}})}  \nonumber \\
{\setbox0=\hbox{\edth} \kern-.025em\copy0\kern-\wd0 \kern.05em\copy0\kern-\wd%
0 \kern-0.025em\raise.0433em\box0 } \mathbf{I}&=& -\frac{\Lambda {Q%
} C{k\!\!\!\!^{-} {}}(1-4\Lambda A^2)}{4{\ (\frac{3}{2}-{k\!\!\!\!^{-} {}})}}%
-\frac{3\Lambda A\overline{P}} {{\ (\frac{3}{2}-{k\!\!\!\!^{-} {}})%
}}\overline{\mathbf{I}}  \nonumber \\
{{\setbox0=\hbox{\edth} \kern-.025em\copy0\kern-\wd0 \kern.05em\copy0\kern-%
\wd0 \kern-0.025em\raise.0433em\box0 }^\prime} \mathbf{I} &=&\frac{3\Lambda C%
{Q}{k\!\!\!\!^{-} {}}}{8P^2 (\frac{3}{2}-{k\!\!\!\!^{-}
{}})}- \frac{3\Lambda A\overline{P}}{{\ (\frac{3}{2}-{%
k\!\!\!\!^{-} {}})}}\mathbf{I}  \nonumber \\
{{\setbox0=\hbox{\thorn} \kern-.025em\copy0\kern-\wd0 \kern.05em\copy0\kern-%
\wd0 \kern-0.025em\raise.0433em\box0 }^\prime} \mathbf{I} &=&{\bf W}  \label{tableI}
\end{eqnarray}
where we have completed the table with some spinor {\bf W} as yet undetermined.
In a similar manner as for previous tables, but this time, using  (\ref{bthornp.o}) and (\ref{bthornp.ob})
we find that
\begin{eqnarray}
{\bf W} = {\frac{%
\overline {P} {Q}^2}{A}} W+\frac{\Lambda {Q} C{%
k\!\!\!\!^{-} {}}(1-4\Lambda A^2)}{4{\ (\frac{3}{2}-{k\!\!\!\!^{-} {}})}}%
\mathbf{I}
\quad -\frac{3\Lambda {Q} C{k\!\!\!\!^{-} {}}}{8P^2 (%
\frac{3}{2}-{k\!\!\!\!^{-} {}})}\overline{\mathbf{I}}- \frac{P}{A}%
\mathbf{I}^2+\frac{3\Lambda A\overline{P}}{{(\frac{3}{2}-{%
k\!\!\!\!^{-} {}})}}\mathbf{I}\overline{\mathbf{I}}  \label{Wdefn}
\end{eqnarray}
where $W$ is a zero-weighted \textit{complex} scalar, as yet undetermined.
Once again we have introduced specific factors alongside the unknown scalar $%
W$ for subsequent simplicity in presentation and to ensure that it is zero-weighted.

We would next like to apply the commutators to $\mathbf{I}$, but to complete
this step we will first need to obtain a complete table for $C$. So we apply the
commutators to $A$ and obtain a partial table for $C$ which we complete as
follows,
\begin{eqnarray}
{\setbox0=\hbox{\thorn}\kern-.025em\copy0\kern-\wd0\kern.05em\copy0\kern-\wd0%
\kern-0.025em\raise.0433em\box0}C &=&-\frac{4}{{Q}(\frac{3}{2}-{%
k\!\!\!\!^{-}{}})}  \nonumber \\
{\setbox0=\hbox{\edth}\kern-.025em\copy0\kern-\wd0\kern.05em\copy0\kern-\wd0%
\kern-0.025em\raise.0433em\box0}C &=&-\frac{P\Lambda AC(5\Lambda A^{2}-2)}{(\frac{3}{2}-{k\!\!\!\!^{-}{}})}+\frac{4}{{Q}(\frac{3}{2}%
-{k\!\!\!\!^{-}{}})}\overline{\mathbf{I}}  \nonumber \\
{{\setbox0=\hbox{\edth}\kern-.025em\copy0\kern-\wd0\kern.05em\copy0\kern-\wd0%
\kern-0.025em\raise.0433em\box0}^{\prime }}C &=&-\frac{\overline{P}%
\Lambda AC(5\Lambda A^{2}-2)}{(\frac{3}{2}-{k\!\!\!\!^{-}{}})}+\frac{4}{{%
Q}(\frac{3}{2}-{k\!\!\!\!^{-}{}})}\mathbf{I}  \nonumber \\
{{\setbox0=\hbox{\thorn}\kern-.025em\copy0\kern-\wd0\kern.05em\copy0\kern-\wd%
0\kern-0.025em\raise.0433em\box0}^{\prime }}C &=&
\mathbf{L}  \label{tableC}
\end{eqnarray}%
where $\mathbf{L}$ is a Hermitian $(1,1)$ type spinor of weight $\{\mathbf{2}%
,\mathbf{2}\}$ determined, from (\ref{thornp.ob}) and (\ref{thornp.o}),  to be:
\begin{eqnarray}
\mathbf{L} &=&\frac{{Q}}{A}L+%
\Bigl(\frac{P\Lambda CA(5\Lambda A^{2}-2)}{(%
\frac{3}{2}-{k\!\!\!\!^{-}{}})}\Bigr)\mathbf{I}+\Bigl(%
\frac{\overline{P}\Lambda CA(5\Lambda A^{2}-2)}{{\ (\frac{3}{2}-{%
k\!\!\!\!^{-}{}})}}\Bigr)\overline{\mathbf{I}}  \nonumber \\
&&-\frac{1}{{Q}}\Bigl(\frac{4}{(\frac{3}{2}-{k\!\!\!\!^{-}{}})}%
\Bigr)\mathbf{I}\overline{\mathbf{I}}  \label{L}
\end{eqnarray}

We have  completed the table, in the same manner as we did for previous tables, with a zero-weighted real scalar $L$, as yet undetermined.

The theory requires that we also apply the commutators to the table for $\mathbf{I%
}$, which yields a partial table for complex  $W$,
\begin{eqnarray}
{\setbox0=\hbox{\thorn} \kern-.025em\copy0\kern-\wd0 \kern.05em\copy0\kern- %
\wd0 \kern-0.025em\raise.0433em\box0 } W &=& \frac{\Lambda C(5\Lambda A^2+4){%
\ {k\!\!\!\!^{-} {}}}^2}{{Q}{\ (\frac{3}{2}-{k\!\!\!\!^{-} {}})}^2}
\nonumber \\
{\setbox0=\hbox{\edth} \kern-.025em\copy0\kern-\wd0 \kern.05em\copy0\kern-%
\wd 0 \kern-0.025em\raise.0433em\box0 } W &=& -2P+\frac{%
\Lambda^2C^2A{{k\!\!\!\!^{-} {}}} ^2(-8\Lambda^2A^4+28\Lambda A^2+7)} {8%
\overline{P}{\ (\frac{3}{2}-{k\!\!\!\!^{-} {}})}^2}-\frac{3\Lambda
A \overline{W}}{2\overline{P} (\frac{3}{2}-{k\!\!\!\!^{-} {}})} -%
\frac{\Lambda AW}{\overline{P}}  \nonumber \\
& & \quad-\frac{\Lambda {k\!\!\!\!^{-} {}} L(1-4\Lambda A^2)}{4\overline{%
P}(\frac{3}{2}-{k\!\!\!\!^{-} {}})} -\frac{ \Lambda C{{%
k\!\!\!\!^{-} {}}}^2(5\Lambda A^2+4)}{{Q}{\ (\frac{3}{2}-{%
k\!\!\!\!^{-} {}})}^2}\overline{\mathbf{I}}  \nonumber \\
{{\setbox0=\hbox{\edth} \kern-.025em\copy0\kern-\wd0 \kern.05em\copy0\kern- %
\wd0 \kern-0.025em\raise.0433em\box0 }^\prime} W &=& \frac{3\Lambda^2C^2A{{%
k\!\!\!\!^{-} {}}}^2(4\Lambda A^2+5)} {8P{(\frac{3}{2}-{%
k\!\!\!\!^{-} {}})}^2}-\frac{\Lambda A{k\!\!\!\!^{-} {}} W}{P (%
\frac{3}{2}-{k\!\!\!\!^{-} {}})} +\frac{3\Lambda L{k\!\!\!\!^{-} {}}}{4%
P(\frac{3}{2}-{k\!\!\!\!^{-} {}})}  \nonumber \\& & \quad
-\frac{\Lambda C {{k\!\!\!\!^{-} {}}}^2(5\Lambda A^2+4)}{{\mathcal{Q%
}}{\ (\frac{3}{2}-{k\!\!\!\!^{-} {}})}^2}\mathbf{I}\   \label{tableW}
\end{eqnarray}

So we have obtained a core element required in our analysis:   a new spinor $%
\mathbf{I}$ which is not parallel to ${\setbox0=\hbox{o}\kern-.025em\copy0%
\kern-\wd0\kern.05em\copy0\kern-\wd0\kern-0.025em\raise.0433em\box0}$. We   have also
constructed its table, and  then we applied the GIF commutators to $%
\mathbf{I}$ in order to
extract further information. Since $\mathbf{I}$ is uniquely defined in terms
of intrinsic elements of the GIF,  we can now transfer these tables into
the GHP formalism and carry out subsequent calculations in the GHP formalism.

\subsection{Transfering to the GHP formalism}

\label{intne3}

We now
identify this spinor ${\mathbf{I}}$ with the second dyad spinor ${\setbox0=\hbox{$\iota$} \kern-.025em\copy0\kern-\wd%
0 \kern.05em\copy0\kern-\wd0 \kern-0.025em\raise.0433em\box0 }$
of the GHP formalism. Then the two tables for the zero weighted $A,C$ can be
immediately translated into the ordinary GHP scalar operators,
\begin{eqnarray}
\hbox{\rm I}\kern-0.32em\raise0.35ex\hbox{\it o}A &=&0  \nonumber
\label{GHPtableA} \\
\hbox{$\partial$\kern-0.25em\raise0.6ex\hbox{\rm\char'40}}A &=&-2P{%
k\!\!\!\!^{-}{}}  \nonumber \\
\hbox{$\partial$\kern-0.25em\raise0.6ex\hbox{\rm\char'40}}^{\prime }A &=&-2%
\overline{P}{k\!\!\!\!^{-}{}}  \nonumber \\
\hbox{\rm I}\kern-0.32em\raise0.35ex\hbox{\it o}^{\prime }A &=&\frac{{%
Q}}{A}C{k\!\!\!\!^{-}{}}^{2}\label{ghptableA}
\end{eqnarray}%
\begin{eqnarray}
\hbox{\rm I}\kern-0.32em\raise0.35ex\hbox{\it o}C &=&-\frac{4}{{Q}(%
\frac{3}{2}-{k\!\!\!\!^{-}{}})}  \nonumber \\
\hbox{$\partial$\kern-0.25em\raise0.6ex\hbox{\rm\char'40}}C &=&-\frac{%
P\Lambda AC(5\Lambda A^{2}-2)}{(\frac{3}{2}-{k\!\!\!\!^{-}{}})}
\nonumber \\
\hbox{$\partial$\kern-0.25em\raise0.6ex\hbox{\rm\char'40}}^{\prime }C &=&-%
\frac{\overline{P}\Lambda AC(5\Lambda A^{2}-2)}{(\frac{3}{2}-{%
k\!\!\!\!^{-}{}})}  \nonumber \\
\hbox{\rm I}\kern-0.32em\raise0.35ex\hbox{\it o}^{\prime }C &=&\frac{{%
Q}}{A}L  \label{GHPtableC}
\end{eqnarray}
 This translation is carried out using (\ref{thornp}), (\ref{edthp}), (\ref{edth}),
(\ref{thorn}), and is especially simple since the operators are acting on scalars.
The table for complex $W$ can also be easily rewritten in GHP operators, but
it will be more convenient to write down two tables for the real and
imaginary parts of $W$ by putting,
\begin{eqnarray}  \label{defnM}
M &=& {\frac{1}{2}} (W + \overline W) - A
\end{eqnarray}
\begin{eqnarray}  \label{defnB}
B &=& {\frac{i}{2}} (W- \overline W)
\end{eqnarray}
which gives,
\begin{eqnarray}
\hbox{\rm I}\kern-0.32em\raise0.35ex\hbox{\it o} M &=& \frac{\Lambda C {{%
k\!\!\!\!^{-} {}}} ^2(5\Lambda A^2 +4)}{{Q}{\ (\frac{3}{2}-{%
k\!\!\!\!^{-} {}})}^2}  \nonumber \\
\hbox{$\partial$\kern-0.25em\raise0.6ex\hbox{\rm\char'40}} M &=& -\frac{%
2\Lambda A^2P{k\!\!\!\!^{-} {}}}{{\ (\frac{3}{2}-{k\!\!\!\!^{-} {}}%
)}}+ \frac{\Lambda^2PAC^2{{k\!\!\!\!^{-} {}}}^3(-\Lambda A^2+\frac{%
11}{2})}{{\ (\frac{3}{2}-{k\!\!\!\!^{-} {}})}^2}+\frac{\Lambda PL{
k\!\!\!\!^{-} {}}^2}{{(\frac{3}{2}-{k\!\!\!\!^{-} {}})}}-\frac{ 3\Lambda A%
PM}{(\frac{3}{2}-{k\!\!\!\!^{-} {}})}  \nonumber \\
& & \quad -\frac{i2{k\Lambda A\!\!\!\!^{-} {}} BP}{ (\frac{3}{2}-{%
k\!\!\!\!^{-} {}})}  \nonumber \\
\hbox{$\partial$\kern-0.25em\raise0.6ex\hbox{\rm\char'40}}^{\prime}M &=& -%
\frac{2\Lambda A^2\overline{P}{k\!\!\!\!^{-} {}}}{{\ (\frac{3}{2}-{%
k\!\!\!\!^{-} {}})}}+ \frac{\Lambda^2\overline{P}AC^2{{%
k\!\!\!\!^{-} {}}}^3(-\Lambda A^2+\frac{11}{2})}{{\ (\frac{3}{2}-{%
k\!\!\!\!^{-} {}})}^2}+\frac{\Lambda \overline{P}L{k\!\!\!\!^{-} {}
}^2}{{(\frac{3}{2}-{k\!\!\!\!^{-} {}})}}   \quad -\frac{ 3\Lambda A\overline{P}M}{(\frac{3}{2}-{%
k\!\!\!\!^{-} {}})}\nonumber \\
& & +\frac{i2{k\Lambda A\!\!\!\!^{-} {}} B\overline{P}}{ (%
\frac{3}{2}-{k\!\!\!\!^{-} {}})} \   \label{partialtableM}
\end{eqnarray}
and
\begin{eqnarray}
\hbox{\rm I}\kern-0.32em\raise0.35ex\hbox{\it o} B &=& 0  \nonumber \\
\hbox{$\partial$\kern-0.25em\raise0.6ex\hbox{\rm\char'40}} B &=& i\Bigl(-2%
P{k\!\!\!\!^{-} {}}- {\Lambda^2PAC^2{{k\!\!\!\!^{-} {}} }%
^2}-\Lambda PL{k\!\!\!\!^{-} {}}-2\Lambda APM\Bigr)
\nonumber \\
\hbox{$\partial$\kern-0.25em\raise0.6ex\hbox{\rm\char'40}}^{\prime}B &=&
-i\Bigl(-2\overline{P}{k\!\!\!\!^{-} {}}- {\Lambda^2\overline{%
P}AC^2{{k\!\!\!\!^{-} {}} }^2}-\Lambda \overline{P}L{%
k\!\!\!\!^{-} {}}-2\Lambda A\overline{P}M\Bigr)
\label{GHPtableB}
\end{eqnarray}
In the table (\ref{tableI}) for $\mathbf{I}$ we will now  make the substitution $W=A+M-iB$: this table is needed to  calculate $\tau' ,\rho' ,\sigma' ,\kappa' $ at the end of this subsection.  However, we shall have no further need of this table in constructing the metric since it only deals with the choice of direction for the second spinor; on the other hand, we will use the table when we consider the Karlhede classification.

From (\ref{tablePQ}) and (\ref{thornp}), (\ref{edthp}), (\ref{edth}),
(\ref{thorn}), the GHP tables for the weighted scalars $P,{%
Q},$ are
\begin{eqnarray}
\hbox{\rm I}\kern-0.32em\raise0.35ex\hbox{\it o} P &=& 0  \nonumber
\\
\hbox{$\partial$\kern-0.25em\raise0.6ex\hbox{\rm\char'40}} {P} &=&
-2P^2\Lambda A  \nonumber \\
\hbox{$\partial$\kern-0.25em\raise0.6ex\hbox{\rm\char'40}}^{\prime}{\mathcal{%
P}} &=& \Lambda A  \nonumber \\
\hbox{\rm I}\kern-0.32em\raise0.35ex\hbox{\it o}^{\prime}{P} &=& 0
\label{GHPtableP}
\end{eqnarray}
\begin{eqnarray}
\hbox{\rm I}\kern-0.32em\raise0.35ex\hbox{\it o} {Q} &=& 0
\nonumber \\
\hbox{$\partial$\kern-0.25em\raise0.6ex\hbox{\rm\char'40}} {Q} &=&-%
{Q}P\Lambda A  \nonumber \\
\hbox{$\partial$\kern-0.25em\raise0.6ex\hbox{\rm\char'40}}^{\prime}{\mathcal{%
Q}} &=&-{Q}\overline {P}\Lambda A  \nonumber \\
\hbox{\rm I}\kern-0.32em\raise0.35ex\hbox{\it o}^{\prime}{{Q}} &=&
0  \label{GHPtableQ}
\end{eqnarray}

\smallskip

We began with GIF tables (\ref{tableA}), (\ref{tablePQ}), for the
zero-weighted scalar $A$ and the weighted complex scalar $(\overline{\mathcal{%
P}}Q)$, from which the GIF commutators generated GIF tables (\ref%
{tableC}), (\ref{tableI}), for the zero-weighted scalar $C$ and the spinor $%
\mathbf{I}$ respectively; furthermore, the GIF commutators acting on the GIF
table for $\mathbf{I}$ generated the partial GIF table (\ref{tableW}), for
the complex zero-weighted $W$. We have now given the equivalent GHP tables (\ref{GHPtableP}), (\ref%
{GHPtableQ}), and (\ref{GHPtableA}), (\ref%
{GHPtableC}), respectively
for the weighted scalars $P,Q$  and the zero-weighted scalars $A,C$  as well as the partial GHP tables  (\ref{partialtableM}), (\ref{GHPtableB}) for the zero-weighted
scalars $M,B$.

In addition to extracting information by applying the GIF commutators to the
table for  $\mathbf{I}$, the theory requires that we obtain complete tables
for four real zero-weighted scalars (coordinate candidates) and one complex
weighted scalar, and apply the commutators to all five of these scalars.
Already we have applied the GIF commutators to the table for the complex
weighted scalar $\overline{P}Q$; \ the zero-weighted
scalars $A,C,M,B$ suggest themselves as the four coordinate candidates, and
hence we will need to ensure that the commutators are applied to all four,
to extract all the information.

Then, \textit{providing that these scalars are functionally independent},
they can be adopted as coordinates. It will be easier to check for this
functional independence after we have simplified the structure of the tables
and after we have also completed the calculation by applying the commutators
to all four candidates.

\smallskip

For subsequent calculations we will require the GHP commutators, which in
turn require the missing four GHP spin coefficients. These four GHP spin
coefficients follow immediately from  (\ref{rho'}) and the
table for $\mathbf{I}$ (\ref{tableI}), and are given by
\begin{eqnarray}  \label{spincs}
\tau^{\prime}= - {\setbox0=\hbox{$\iota$} \kern-.025em\copy0\kern-\wd0 \kern%
.05em\copy0\kern-\wd0 \kern-0.025em\raise.0433em\box0 }^B {\setbox0=%
\hbox{$\iota$} \kern-.025em\copy0\kern-\wd0 \kern.05em\copy0\kern-\wd0 \kern%
-0.025em\raise.0433em\box0 }^C\bar{\i}^{C^{\prime}} {\setbox0=\hbox{\thorn} %
\kern-.025em\copy0\kern-\wd0 \kern.05em\copy0\kern-\wd0 \kern-0.025em\raise%
.0433em\box0 }_{CC^{\prime}}\, ({\setbox0=\hbox{$\iota$} \kern-.025em\copy0%
\kern-\wd0 \kern.05em\copy0\kern-\wd0 \kern-0.025em\raise.0433em\box0 }_B) =%
\frac{3\Lambda A\overline{P}}{(\frac{3}{2}-{k\!\!\!\!^{-} {}})} \nonumber \\
\rho^{\prime}= -{\setbox0=\hbox{$\iota$} \kern-.025em\copy0\kern-\wd0 \kern%
.05em\copy0\kern-\wd0 \kern-0.025em\raise.0433em\box0 }^B {\setbox0=%
\hbox{$\iota$} \kern-.025em\copy0\kern-\wd0 \kern.05em\copy0\kern-\wd0 \kern%
-0.025em\raise.0433em\box0 }^C\bar{\i}^{C^{\prime}}\bar{\i}^{D^{\prime}} {%
\setbox0=\hbox{\edth} \kern-.025em\copy0\kern-\wd0 \kern.05em\copy0\kern-\wd%
0 \kern-0.025em\raise.0433em\box0 }_{CC^{\prime}D^{\prime}}\, ({\setbox0=%
\hbox{$\iota$} \kern-.025em\copy0\kern-\wd0 \kern.05em\copy0\kern-\wd0 \kern%
-0.025em\raise.0433em\box0 }_B) = \frac{\Lambda {Q} C{%
k\!\!\!\!^{-} {}}(1-4\Lambda A^2)}{4{\ (\frac{3}{2}-{k\!\!\!\!^{-} {}})}}
\nonumber \\
\sigma^{\prime}= -{\setbox0=\hbox{$\iota$} \kern-.025em\copy0\kern-\wd0 \kern%
.05em\copy0\kern-\wd0 \kern-0.025em\raise.0433em\box0 }^B {\setbox0=%
\hbox{$\iota$} \kern-.025em\copy0\kern-\wd0 \kern.05em\copy0\kern-\wd0 \kern%
-0.025em\raise.0433em\box0 }^C{\setbox0=\hbox{$\iota$} \kern-.025em\copy0%
\kern-\wd0 \kern.05em\copy0\kern-\wd0 \kern-0.025em\raise.0433em\box0 }^D%
\bar{\i}^{C^{\prime}} {\setbox0=\hbox{\edth} \kern-.025em\copy0\kern-\wd0 %
\kern.05em\copy0\kern-\wd0 \kern-0.025em\raise.0433em\box0 }^{\prime}_{CDD^{\prime}}
\, ({\setbox0=\hbox{$\iota$} \kern-.025em\copy0\kern-\wd0 \kern.05em\copy0%
\kern-\wd0 \kern-0.025em\raise.0433em\box0 }_B) = -\frac{3\Lambda C{\mathcal{Q%
}}{k\!\!\!\!^{-} {}}}{8P^2 (\frac{3}{2}-{k\!\!\!\!^{-} {}})}
\nonumber \\
\kappa^{\prime}= - {\setbox0=\hbox{$\iota$} \kern-.025em\copy0\kern-\wd0 %
\kern.05em\copy0\kern-\wd0 \kern-0.025em\raise.0433em\box0 }^B {\setbox0=%
\hbox{$\iota$} \kern-.025em\copy0\kern-\wd0 \kern.05em\copy0\kern-\wd0 \kern%
-0.025em\raise.0433em\box0 }^C{\setbox0=\hbox{$\iota$} \kern-.025em\copy0%
\kern-\wd0 \kern.05em\copy0\kern-\wd0 \kern-0.025em\raise.0433em\box0 }^D%
\bar{\i}^{C^{\prime}}\bar{\i}^{D^{\prime}} {{\setbox0=\hbox{\thorn} \kern%
-.025em\copy0\kern-\wd0 \kern.05em\copy0\kern-\wd0 \kern-0.025em\raise.0433em%
\box0 }^\prime}_{CDC^{\prime}D^{\prime}}\, ({\setbox0=\hbox{$\iota$} \kern%
-.025em\copy0\kern-\wd0 \kern.05em\copy0\kern-\wd0 \kern-0.025em\raise.0433em%
\box0 }_B) ={\frac{\overline {P} {Q}^2}{A}} (A+M-iB)
\end{eqnarray}
and of course $\tau=P /A$; these should now be substituted into the GHP
commutators (\ref{GHPcomm}).

\subsection{Simplifying and completing the tables in the GHP operators}

\label{intne4}

We have already obtained a GHP table (\ref{ghptableA}) for the real zero-weighted scalar $A
$, and in addition applied the commutators, which is required for a
coordinate candidate.

From this, we have also obtained a GHP table (\ref{GHPtableC}) for the real
zero-weighted scalar $C$, and when we apply the GHP commutators we obtain
the partial table for the real zero-weighted scalar $L$,
\begin{eqnarray}
\hbox{\rm I}\kern-0.32em\raise0.35ex\hbox{\it o}L &=&\frac{-18\Lambda
^{2}CA^{3}{{k\!\!\!\!^{-}{}}}}{{Q}{\ (\frac{3}{2}-{k\!\!\!\!^{-}{}}%
)}^{2}}  \nonumber \\
\hbox{$\partial$\kern-0.25em\raise0.6ex\hbox{\rm\char'40}}L &=&-\frac{%
\Lambda C^{2}P{{k\!\!\!\!^{-}{}}}^{2}(11\Lambda A^{2}-2)}{{\ (%
\frac{3}{2}-{k\!\!\!\!^{-}{}})}^{2}}-\frac{\Lambda LAP(4\Lambda
A^{2}-1)}{{\ (\frac{3}{2}-{k\!\!\!\!^{-}{}})}}+\frac{4P}{{\ (\frac{%
3}{2}-{k\!\!\!\!^{-}{}})}}(A+M+iB)  \nonumber \\
\hbox{$\partial$\kern-0.25em\raise0.6ex\hbox{\rm\char'40}}^{\prime }L &=&-%
\frac{\Lambda C^{2}\overline{P}{{k\!\!\!\!^{-}{}}}^{2}(11\Lambda
A^{2}-2)}{{\ (\frac{3}{2}-{k\!\!\!\!^{-}{}})}^{2}}-\frac{\Lambda LA\overline{%
P}(4\Lambda A^{2}-1)}{{\ (\frac{3}{2}-{k\!\!\!\!^{-}{}})}}\nonumber \\
& & +\frac{4%
\overline{P}}{{\ (\frac{3}{2}-{k\!\!\!\!^{-}{}})}}(A+M-iB)
\label{partialtableL}
\end{eqnarray}%
So we can adopt $C$ as a second coordinate candidate and add the partial table for $%
L$ to our equations.

We would next like to complete the partial tables for two of $M,B,L$,
and then apply the commutators to each to exploit them as two more coordinate candidates. However, it is clear from the complicated partial tables above that these calculations would  be long, and it
will be easier if we first do a little rearranging and
relabelling. The simpler the form  which we can obtain for our tables for the four coordinate candidates, the simpler the form will be for the associated metric.

A direct substitution of $M$ by $T$ via
\begin{equation}
T = -\frac{1}{\sqrt{2}{{k\!\!\!\!^{-} {}}}^
\frac{1}{2}}M-\frac{\Lambda {{k\!\!\!\!^{-} {}}}^
\frac{3}{2}(9{k\!\!\!\!^{-} {}}+4)}{8\sqrt{2}} C^2-\frac{\Lambda
A{k\!\!\!\!^{-} {}}^ \frac{1}{2}}{2\sqrt{2}}L \label{T=M}
\end{equation}
enables the complicated partial table for $M$ to be replaced with a very
simple table for $T$,
\begin{eqnarray}
\hbox{\rm I}\kern-0.32em\raise0.35ex\hbox{\it o} T &=& 0  \nonumber \\
\hbox{$\partial$\kern-0.25em\raise0.6ex\hbox{\rm\char'40}} T &=& 0  \nonumber
\\
\hbox{$\partial$\kern-0.25em\raise0.6ex\hbox{\rm\char'40}}^{\prime}T &=& 0
\nonumber \\
\hbox{\rm I}\kern-0.32em\raise0.35ex\hbox{\it o}^{\prime}T &=&\frac{{%
Q}{{k\!\!\!\!^{-} {}}}^{ \frac{1}{4}}}{A} F  \label{GHPtableT}
\end{eqnarray}
which we have completed in the usual way for the, as yet undetermined,
zero-weighted scalar function $F$. (Once again, the particular choice of
factors multiplying the unknown function $F$ is simply to ensure that $F$ is a zero-weighted scalar, and to shorten the
presentation of the details of the subsequent calculations.) So we decide to replace $M$ with $T$ as a third coordinate candidate.

It now remains to get a simpler replacement for the rather complicated tables (\ref{GHPtableB})  and  (\ref{partialtableL}), for $B$ and  $L$ respectively.

Making  a direct substitution of $L$ with $S$ in (\ref{GHPtableB})   via
\begin{eqnarray}
S= (2
{k\!\!\!\!^{-} {}}+ {\Lambda^2AC^2{{k\!\!\!\!^{-} {}} }%
^2}+\Lambda L{k\!\!\!\!^{-} {}}+2\Lambda AM)/\Lambda {k\!\!\!\!^{-} {}}^{1/2}
\label{S=L}
\end{eqnarray}
gives the simpler form
\begin{eqnarray}
\hbox{\rm I}\kern-0.32em\raise0.35ex\hbox{\it o} B &=& 0  \nonumber \\
\hbox{$\partial$\kern-0.25em\raise0.6ex\hbox{\rm\char'40}} B &=&- iP\Lambda {k\!\!\!\!^{-} {}}^{1/2}S\nonumber \\
\hbox{$\partial$\kern-0.25em\raise0.6ex\hbox{\rm\char'40}}^{\prime}B &=&
iP\Lambda {k\!\!\!\!^{-} {}}^{1/2}S
\label{GHPtableB.1}
\end{eqnarray}
as well as replacing the complicated partial table for $L$ with the  simpler partial table for $S$
\begin{eqnarray}
\hbox{\rm I}\kern-0.32em\raise0.35ex\hbox{\it o} S &=& 0  \nonumber \\
\hbox{$\partial$\kern-0.25em\raise0.6ex\hbox{\rm\char'40}}{S} &=&4i \mathcal{%
P}{{k\!\!\!\!^{-} {}}}^{\frac{1}{2}}B  \nonumber \\
\hbox{$\partial$\kern-0.25em\raise0.6ex\hbox{\rm\char'40}}^{\prime}{S} &=&-4i%
\overline {P}{{k\!\!\!\!^{-} {}}}^{\frac{1 }{2}}B
\label{GHPtableS}
\end{eqnarray}
The term in $L$ in the table (\ref{GHPtableC}) for $C$ will now be replaced, using (\ref{S=L}) and  (\ref{T=M}), with
\begin{eqnarray}
L  &=& \Bigl(S+{2\sqrt{2}A  }T  -\frac{2 {{{k\!\!\!\!^{-} {}}}^{1/2}} }{\Lambda} +
\frac {9\Lambda^2 A^3 C^2 {k\!\!\!\!^{-} {}}^{3/2}}{4}  \Bigr)/ {{{k\!\!\!\!^{-} {}}}^{1/2}}(\frac{3}{2}- {k\!\!\!\!^{-} {}})
\label{Sdefn}
\end{eqnarray}

These two partial tables (\ref{GHPtableB.1}) and (\ref{GHPtableS})  are much simpler in appearance than (\ref{GHPtableB}) and (\ref{partialtableL}), and our next step would seem to be to choose $B$ or $S$ as the fourth coordinate candidate and complete its table in the usual manner, and then apply the commutators to it; the other scalar function would then be defined by its partial table. Unfortunately, because of the coupled nature of  $B$ or $S$ in the two tables (\ref{GHPtableB.1}) and (\ref{GHPtableS}) the subsequent application of the commutators to such an arrangement gets very complicated; therefore it is more convenient to make one more rearrangement.

So we make a substitution of $B$ with $V$ by
\begin{eqnarray}
V=2B/S  \label{Vdefn}
\end{eqnarray}
Clearly this substitution is not valid when  $S=0$; so we shall assume $S\ne 0$ in the remainder of this section, and we shall later have to look at the special case $S=0$ separately.

This means
 that now a comparatively simple  table  for ${V}$ replaces the partial table (\ref{GHPtableB.1}) for $B$,
\begin{eqnarray}
\hbox{\rm I}\kern-0.32em\raise0.35ex\hbox{\it o} {V} &=& 0  \nonumber \\
\hbox{$\partial$\kern-0.25em\raise0.6ex\hbox{\rm\char'40}} {V} &=& -2i%
P{{k\!\!\!\!^{-} {}}}^\frac{1}{2} ({V}^2+{\Lambda})  \nonumber \\
\hbox{$\partial$\kern-0.25em\raise0.6ex\hbox{\rm\char'40}}^{\prime}{V} &=&2i%
\overline {P}{{k\!\!\!\!^{-} {}}}^ \frac{1}{2}({V}^2+{\Lambda})
\nonumber \\
\hbox{\rm I}\kern-0.32em\raise0.35ex\hbox{\it o}^{\prime}{V} &=& \frac{{%
Q}{{k\!\!\!\!^{-} {}}}^{ \frac{1}{4}}}{2A}({V}^2+{\Lambda}) H
\label{GHPtableV}
\end{eqnarray}
which we have completed in the usual way with the, as yet undetermined,
zero-weighted scalar function $H$.

In order to obtain a still simpler form for this table, and a corresponding simpler form for the metric, we can now divide across the whole table by $({V}^2+{\Lambda})$  and by integration define an alternative coordinate
candidate to ${V}$ with a simpler table.

However, it is important to note that in order to integrate with respect to $V$ we have made the assumption that $V \ne$ constant; this assumption also ensures that $V^2+\Lambda \ne 0$.  Hence we will need to consider separately $V=$ constant as a special case.

So we define
\begin{equation}
X = \int \frac {d {V}}{{\ V}^2+\Lambda} = \frac{1}{\sqrt{|\Lambda|}} \hbox{tan[h]}^{-1}(\frac{V}{\sqrt{%
|\Lambda|}})  \label{definX}
\end{equation}
where we have introduced this compact notation
\begin{eqnarray}  \label{tanh-1}
 \hbox{tan[h]}^{-1}(\frac{V}{\sqrt{
|\Lambda|}}) =\left\{\matrix{- \tanh^{-1}(\frac{V}{\sqrt{-
\Lambda}}) \quad \hbox{for} \quad\Lambda < 0 \cr
   \tan^{-1}(\frac{V}{\sqrt{\Lambda}})\quad
\hbox{for} \quad \Lambda > 0} \right.
\end{eqnarray}
and we have now the table
\begin{eqnarray}
\hbox{\rm I}\kern-0.32em\raise0.35ex\hbox{\it o} {X} &=& 0  \nonumber \\
\hbox{$\partial$\kern-0.25em\raise0.6ex\hbox{\rm\char'40}} {X} &=& -2i%
P{{k\!\!\!\!^{-} {}}}^\frac{1}{2}  \nonumber \\
\hbox{$\partial$\kern-0.25em\raise0.6ex\hbox{\rm\char'40}}^{\prime}{X} &=&2i%
\overline {P}{{k\!\!\!\!^{-} {}}}^ \frac{1}{2}  \nonumber \\
\hbox{\rm I}\kern-0.32em\raise0.35ex\hbox{\it o}^{\prime}{X} &=& \frac{{%
Q}{{k\!\!\!\!^{-} {}}}^{ \frac{1}{4}}}{2A} H \quad
\label{GHPtableX}
\end{eqnarray}

Since this table turns out to be more manageable, we will adopt ${X}$ as the fourth
coordinate candidate.

The partial table for $S$ is now modified to
\begin{eqnarray}
\hbox{\rm I}\kern-0.32em\raise0.35ex\hbox{\it o} S &=& 0  \nonumber \\
\hbox{$\partial$\kern-0.25em\raise0.6ex\hbox{\rm\char'40}}{S} &=&2i\,
P{{k\!\!\!\!^{-} {}}}^{\frac{1}{2}} S \sqrt{ |\Lambda|}\,\hbox{tan[h]}(\sqrt{ |\Lambda|}\, X)  \nonumber
\\
\hbox{$\partial$\kern-0.25em\raise0.6ex\hbox{\rm\char'40}}^{\prime}{S}
&=&-2i \overline {P}{{k\!\!\!\!^{-} {}}}^{\frac{1}{2}} S \sqrt{ |\Lambda|}\,\hbox{tan[h]}(\sqrt{ |\Lambda|}\, X) .  \label{GHPtableS2}
\end{eqnarray}
where
\begin{eqnarray}
\hbox{tan[h]}(\sqrt{ |\Lambda|}\, x)  =\left\{\matrix{  - \tanh(\sqrt{ -\Lambda}\,x) \quad \hbox{for} \quad\Lambda < 0  \cr
  \tan (\sqrt{ \Lambda}\,x)\quad %
\hbox{for} \quad \Lambda > 0
 \label{tanh}}\right.
\end{eqnarray}

Earlier, we postponed applying the GIF commutators to the two real scalars $B,M$, so we need to apply the GHP commutators equivalently to their replacements, the two real
zero-weighted scalars $T, X$. Applying the GHP commutators (\ref{GHPcomm}) to (\ref{GHPtableT}) and (\ref{GHPtableX}) gives the simple partial tables
for $F$ and $H$ respectively,
\begin{eqnarray}
\hbox{\rm I}\kern-0.32em\raise0.35ex\hbox{\it o}F &=&0  \nonumber \\
\hbox{$\partial$\kern-0.25em\raise0.6ex\hbox{\rm\char'40}}F &=&0  \nonumber
\\
\hbox{$\partial$\kern-0.25em\raise0.6ex\hbox{\rm\char'40}}^{\prime }F &=&0
\label{GHPtableF}
\end{eqnarray}

\begin{eqnarray}
\hbox{\rm I}\kern-0.32em\raise0.35ex\hbox{\it o} H &=& 0  \nonumber \\
\hbox{$\partial$\kern-0.25em\raise0.6ex\hbox{\rm\char'40}}{H} &=& 0
\nonumber \\
\hbox{$\partial$\kern-0.25em\raise0.6ex\hbox{\rm\char'40}}^{\prime}{H} &=& 0
\label{GHPtableH}
\end{eqnarray}

The rather extensive relabelling and rearranging --- to obtain the two tables (\ref{GHPtableT}) and (\ref{GHPtableX}) and the constraints (\ref{GHPtableS2}), (\ref{GHPtableF}) and (\ref{GHPtableH}) --- which we have just carried out was in order
to obtain such simple and manageable forms. Clearly the gradient
vector $\nabla F$ is parallel to $\nabla T$; this means that the scalar function $F$
is an arbitrary function of only the one coordinate candidate $T$ (and
independent of the other coordinate candidates ${X}, C, A$). Similarly, from (\ref{GHPtableH})  the
function $H$ is also an arbitrary function of only the one coordinate
candidate $T$. The function $S$ in (\ref{GHPtableS2}).  has a more complicated structure; we shall
find it as the solution of a partial differential equation when we translate
into explicit coordinates.

Since we have applied the commutators to $\mathbf{I}$ and to $P$ and $Q$, as well as to $A,C,T,X,$ we have obtained, in an
explicit form, \textit{all} the information about this class of spaces. So
we have completed the formal integration procedure for these spaces; all the
information has been extracted in the generic case, by which we mean the
case where we have assumed that \textit{the four zero-weighted real scalar
functions, $A,C,T,X$ are functionally independent}; these are our
coordinate candidates which we intend to adopt as coordinates.

In summary, we note that we have complete tables (\ref{GHPtableA}), (\ref%
{GHPtableC}), (\ref{GHPtableT}),  (\ref{GHPtableX}),  for the four
zero-weighted real scalar functions, $A,C,T,X$ respectively; $L$ in (\ref{GHPtableC}) is replaced by $S$ from  (\ref{Sdefn}). Clearly our tables for the zero-weighted scalars
$A,C,T,X$ and for the weighted scalars $P$ and $Q$
are not complete and involutive \textit{by themselves}, since they contain
also the zero-weighted scalar functions $S,H,F$. However, by applying the
commutators to these four scalars we have obtained the constraint equations
in the form of the partial tables (\ref{GHPtableS2}), (\ref{GHPtableF}), (%
\ref{GHPtableH}) for these additional scalar functions, which, taken together with
the tables (\ref{GHPtableA}), (\ref{GHPtableC}), (\ref{GHPtableT}), (\ref%
{GHPtableX}), (\ref{GHPtableP}), (\ref{GHPtableQ}) supply
a complete and involutive system.

In the remainder of this section we will obtain the coordinate version of
the tetrad vectors, and hence  the metric.

As we emphasised in the last subsection, before we can adopt the coordinate
candidates as coordinates, we must confirm that they are functionally
independent. First of all we check on the possibility of these four scalars
being constant: since we are assuming in this section that ${k\!\!\!\!^{-} {}%
}\ne 0$, then none of $A,C,X$ can be constant, but $T$ may be. From the
tables if follows that $T$ is constant iff $F=0$. Hence, in this section,
the additional assumption that $F \ne 0 $ is sufficient to ensure that none
of the coordinate candidates are constant. Moreover, when we assume that
none of the coordinate candidates are constant, a check of the determinant
formed from their four tables (\ref{GHPtableT}), (\ref{GHPtableC}), (\ref%
{GHPtableA}), (\ref{GHPtableX}), confirms that the four coordinate
candidates are indeed functionally independent --- providing $F \ne 0 $.
Hence we will complete this section for the generic case with the additional
assumption $F\ne 0 $ ensuring that the coordinate candidates $A,C,T,{X}$ can
be adopted as explicit coordinates.

\smallskip In addition, we must not forget that in order that $X$ could be a
coordinate candidate, we made the additional assumptions that
 $V\neq $ constant, and  $S\ne 0$.
In Section 6 we will  look separately at the special case $V = $ constant, and in Section 7 we will investigate the special cases with $F=0=S$.

\subsection{Using coordinate candidates as coordinates}

\label{intne5}

If we now make the obvious choice of the coordinate candidates as coordinates

\begin{equation}
t=T,\quad c=C, \quad a=A, \quad x={X}  \label{coord}
\end{equation}
the above four tables for the zero-weighted scalars enable us to immediately write down
the tetrad vectors in the coordinates $t,c,a,{x}$,
\begin{eqnarray}
l^i &=& \frac{1}{{Q}} \Bigl(0, \ -\frac{4}{(\frac{3}{2}-{k\!\!\!\!^{-} {}})}, \ 0, \
0\Bigr)  \nonumber \\
m^i &=& P\Bigl(0, \ -\frac{\Lambda(5\Lambda a^2-2)ac}{{{(\frac{3}{2}-{k\!\!\!\!^{-} {}})}} }, \ -2{{k\!\!\!\!^{-} {}}}, \ -2i{{%
k\!\!\!\!^{-} {}}}^\frac{1}{2}\Bigr)  \nonumber \\
\overline m^i &=& \overline {P}\Bigl(0, \ -\frac{\Lambda(5\Lambda a^2-2)ac}{{{(\frac{3}{2}-{k\!\!\!\!^{-} {}})}} }, \ -2{{k\!\!\!\!^{-} {}%
}}, \ 2i{{k\!\!\!\!^{-} {}}}^\frac{1}{2}\Bigr)  \nonumber \\
n^i &=& \frac{{Q}}{a}\Bigl( F{{k\!\!\!\!^{-} {}}}^{\frac{1}{4}}, \ L, \ {{k\!\!\!\!^{-} {}}}^2 c, \ \frac{H{k\!\!\!\!^{-} {}}^{\frac{1}{4}}}{2}\Bigr)
\label{frame1}
\end{eqnarray}
where the function $L$ is given in terms of $S$ by (\ref{Sdefn}), the functions $S, \ H, \
F $ are respectively solutions of the partial tables (\ref{GHPtableS2}), (%
\ref{GHPtableH}), (\ref{GHPtableF}), and now ${k\!\!\!\!^{-} {}}=\Lambda
a^2+1/2$ from (\ref{A1defn}).

As noted in the last section, $F$ and $H$ respectively will be arbitrary
functions of only the one coordinate $t$, so we will write $-4F=\alpha_2(t)$
and $-2H=\alpha_3(t)$ --- subject to the restrictions made in the
calculations in this section that $F\ne 0$ which implies that $%
\alpha_2(t)\ne 0$ (note there is no restriction on $\alpha_3(t)$, which is
a completely arbitrary function of $t$, including the zero function).

The partial table (\ref{GHPtableS2}) for $S$ now becomes, via the tetrad, a
system of partial differential equations in the chosen coordinates,
\[
\frac{\partial S}{\partial c}=0
\]
\begin{equation}
2{k\!\!\!\!^{-} {}}\frac{\partial S}{\partial a} +2i{{k\!\!\!\!^{-} {}}}^{%
\frac{1}{2}}\frac{ \partial S}{\partial {x}} = -2 i{{k\!\!\!\!^{-} {}}}^{%
\frac{1}{2}} S\sqrt{ |\Lambda|}\,\hbox{tan[h]}(\sqrt{ |\Lambda|}\, x)
\label{pdeS}
\end{equation}
which shows that $S$ is independent of the coordinates $c$ and $a$, and we
easily find the solution using (\ref{definX})
\begin{eqnarray}
S(t,{x})=\alpha_1(t)\,\hbox{cos[h]}(\sqrt{ |\Lambda|}\,x)  \label{Ssolution}
\end{eqnarray}
where $\hbox{cos[h]}(\sqrt{ |\Lambda|}\, x)$ is given by
\begin{eqnarray}
\hbox{cos[h]}(\sqrt{ |\Lambda|}\, x)  =\left\{\matrix{  \cosh(\sqrt{ -\Lambda}\,x) \quad \hbox{for} \quad\Lambda < 0
\cr    \cos (\sqrt{ \Lambda}\,x)\quad
\hbox{for} \quad \Lambda > 0}\right.
 \label{cosh}
\end{eqnarray}
and $\alpha_1(t)$ is an
arbitrary function of $t$, excluding the zero function, since we are assuming $S\ne 0$ in this section.

\smallskip

It follows immediately from the equation
\begin{equation}
g^{ij}=2l^{(i}n^{j)}-2m^{(i}{\overline m}^{j)}  \label{metric}
\end{equation}
that the metric $g^{ij}$, in the coordinates $t,c,a,{x}$, is given by
\begin{equation}
g^{ij}= \left(
\begin{array}{lccr}
0 & \frac{{k\!\!\!\!^{-} {}}^{1/4}\alpha_2(t)}{a{(\frac{3}{2}-{k\!\!\!\!^{-} {}})}%
} & 0 & 0 \cr \frac{{k\!\!\!\!^{-} {}}^{1/4}\alpha_2(t)}{ a(\frac{3}{2}-{%
k\!\!\!\!^{-} {}})} & -\frac{8 }{a {{k\!\!\!\!^{-} {}}}^{1/2}(\frac{3}{2}-{%
k\!\!\!\!^{-} {}})^{2}}Z & -\frac{2{{k\!\!\!\!^{-} {}}}(5\Lambda^2 a^4+1)c}{a(\frac{3}{2}-{%
k\!\!\!\!^{-} {}})} &  \frac{{k\!\!\!\!^{-} {}}^{1/4}\alpha_3(t)}{a(\frac{3}{2}-{%
k\!\!\!\!^{-} {}})}\cr 0 & -\frac{2{{k\!\!\!\!^{-} {}}}(5\Lambda^2 a^4+1)c}{a(\frac{3%
}{2}-{k\!\!\!\!^{-} {}})} & -4 {{k\!\!\!\!^{-} {}}}^2 & 0 \cr 0 & \frac{{%
k\!\!\!\!^{-} {}}^{1/4}\alpha_3(t)}{a(\frac{3}{2}-{k\!\!\!\!^{-} {}})} & 0 &
-4{k\!\!\!\!^{-} {}}%
\end{array}
\right)  \label{metric1}
\end{equation}
where ${k\!\!\!\!^{-} {}}=\Lambda a^2 +1/2$, and $Z$ is given in terms of $S$ from (\ref{Sdefn}) by,
\begin{eqnarray}
Z &=& {(\frac{3}{2}-{k\!\!\!\!^{-} {}})}{{k\!\!\!\!^{-} {}}}^{1/2}\Bigl(L  +\frac{\Lambda^2(5\Lambda a^2-2)^2a^3c^2 }{8
{(\frac{3}{2}-{k\!\!\!\!^{-} {}})} }\Bigr) \nonumber \\
& =& S+{2\sqrt{2}a  }t  -\frac{2 {{{k\!\!\!\!^{-} {}}}^{1/2}} }{\Lambda}
 + \frac {9\Lambda^2 a^3 c^2 {k\!\!\!\!^{-} {}}^{3/2}}{4}
+\frac{{{k\!\!\!\!^{-} {}}}^{1/2}\Lambda^2(5\Lambda a^2-2)^2a^3c^2 }{8
} \nonumber \\
&=&  \alpha_1(t)\hbox{cos[h]}(\sqrt{|\Lambda|}x) +{2\sqrt{2}a  }t  -\frac{2 {{{k\!\!\!\!^{-} {}}}^{1/2}} }{\Lambda}
\nonumber \\  & & \qquad \qquad\qquad\qquad +\frac {\Lambda^2 {{k\!\!\!\!^{-} {}}}^{1/2} a^3 c^2 (25\Lambda^2a^4-2\Lambda a^2+13)} {8}
  \label{Zcoords1}
\end{eqnarray}
where $\hbox{cos[h]}(\sqrt{|\Lambda|}x) $ is given by (\ref{cosh}), and  $\alpha _{3}(t)$ is completely arbitrary.
\smallskip

We must remember that, in order to justify taking $t$ as a coordinate, we
have assumed that $\alpha _{2}(t)\neq 0$, and in order to justify taking $x$ as a coordinate, we
have assumed that $\alpha _{1}(t)\neq 0$; furthermore we have assumed at
certain stages in our calculations that $V\neq $ constant. So this metric is not necessarily the most general form for this
class of spacetimes.

In the following sections we will first look at the
excluded cases separately, and then obtain a more general form of the metric which will
include all  such previously excluded cases.

\section{The integration procedure for $\Lambda + \protect\tau\overline
\protect\tau \ne 0$: special case $V =\hbox{ constant}$, and combined case}

\label{special}

\subsection{The special case with $V= $ constant\label{genne1}}

When we substitute the condition $V =$ constant into (\ref
{GHPtableV})  we find that this case can only
occur for a \textit{negative cosmological constant}. So if we write
\[
\lambda =\pm \sqrt{-\Lambda }
\]%
then we find $V=\lambda $. The calculations in Section \ref{intne} up to (%
\ref{GHPtableS}) are still valid. Since neither ${X}$, nor constant $V$, can
 be a coordinate as in the last section, we must find a
replacement coordinate candidate which is functionally independent of the
other three $A,C,T$. We shall continue to assume in this section that $F\neq
0\ne S$.

Substitution of $V=\lambda $ into (\ref{Vdefn}) modifies (\ref{GHPtableS}) to give the
table for $B$ for this special case,
\begin{eqnarray}
\hbox{\rm I}\kern-0.32em\raise0.35ex\hbox{\it o}B &=&0  \nonumber
\label{GHPtableB1} \\
\hbox{$\partial$\kern-0.25em\raise0.6ex\hbox{\rm\char'40}}B &=&2i{\mathcal{P%
}}{{k\!\!\!\!^{-}{}}}^{\frac{1}{2}} \lambda B  \nonumber \\
\hbox{$\partial$\kern-0.25em\raise0.6ex\hbox{\rm\char'40}}^{\prime }B &=&-2i{%
\overline{P}}{{k\!\!\!\!^{-}{}}}^{\frac{1}{2}} \lambda   B \nonumber
\\
\hbox{\rm I}\kern-0.32em\raise0.35ex\hbox{\it o}^{\prime }B &=&-\frac{{%
Q}}{2A}{{k\!\!\!\!^{-}{}}}^{\frac{1}{4}}    \lambda B G
\end{eqnarray}%
The real zero-weighted scalar $G$ --- as yet undetermined --- has been
chosen to complete the table in the usual manner.

This comparatively simple table suggests $B$ as the replacement coordinate
candidate; this of course will require that $B \ne$ constant, but from (\ref{GHPtableB1}) we then see that the only possible constant value is $B=0$. However, from (\ref{GHPtableB.1}) it follows that $S=0$, and this special class has been excluded from this section.

But an even simpler table is obtained by the substitution
\begin{eqnarray}
e^{-\lambda Y} = |B|
\label{Y=lnB}\end{eqnarray}
giving
\begin{eqnarray}
\hbox{\rm I}\kern-0.32em\raise0.35ex\hbox{\it o}Y &=&0  \nonumber
\label{GHPtableY} \\
\hbox{$\partial$\kern-0.25em\raise0.6ex\hbox{\rm\char'40}}Y &=&-2i{\mathcal{P%
}}{{k\!\!\!\!^{-}{}}}^{\frac{1}{2}}  \nonumber \\
\hbox{$\partial$\kern-0.25em\raise0.6ex\hbox{\rm\char'40}}^{\prime }Y &=&2i{%
\overline{P}}{{k\!\!\!\!^{-}{}}}^{\frac{1}{2}}  \nonumber \\
\hbox{\rm I}\kern-0.32em\raise0.35ex\hbox{\it o}^{\prime }Y &=&\frac{{%
Q}}{2A}{{k\!\!\!\!^{-}{}}}^{\frac{1}{4}}G
\end{eqnarray}%
So preferring $Y$ as our fourth coordinate candidate, we apply the
commutators to get
\begin{eqnarray}
\hbox{\rm I}\kern-0.32em\raise0.35ex\hbox{\it o}G &=&0  \nonumber \\
\hbox{$\partial$\kern-0.25em\raise0.6ex\hbox{\rm\char'40}}{G} &=&0  \nonumber
\\
\hbox{$\partial$\kern-0.25em\raise0.6ex\hbox{\rm\char'40}}^{\prime }{G} &=&0
\label{sGHPtableG}
\end{eqnarray}

The tables (\ref{GHPtableA}), (\ref{GHPtableT}), (\ref{GHPtableC})
respectively for the other three coordinate candidates $A,T,C$ and the
partial table (\ref{GHPtableF}) for the function $F$, are unchanged. $L$
is replaced in (\ref{GHPtableC}) by $S$ from (\ref{Sdefn}), which in return is replaced by $Y$ from
\[
S=2B/\lambda =\frac{2}{\lambda }e^{-\lambda Y}
\]%
from (\ref{Y=lnB}) (remembering there is a $\pm $ included in our definition
of $\lambda $).

We have already noted that $A$ and $C$ cannot be constants, and although $T$ may be, we are  excluding that possibility in this section (since $F\ne 0$); furthermore, it is clear that $Y$ cannot be constant (remembering ${k\!\!\!\!^{-}{}}\neq
0\neq \lambda $). Moreover, an examination of the determinant of the four
tables (\ref{GHPtableA}), (\ref{GHPtableC}), (\ref{GHPtableT}) and (\ref%
{GHPtableY}) shows that the four scalars $A$, $C$, $T$ and $Y$ are
functionally independent and therefore can be chosen as coordinates.

So we now make the obvious choice of the coordinate candidates as
coordinates,
\[
t=T,\qquad c=C,\qquad a=A,\qquad y=Y\ .\label{coord3ss}
\]

We can write down the tetrad vectors in these coordinates by means of the
tables (\ref{GHPtableA}), (\ref{GHPtableC}), (\ref{GHPtableT}) and (\ref%
{GHPtableY}),
\begin{eqnarray}
l^i &=& \frac{1}{{Q}} \Bigl(0, \ -\frac{4}{(\frac{3}{2}-{k\!\!\!\!^{-} {}})}, \ 0, \
0\Bigr)  \nonumber \\
m^i &=& P\Bigl(0, \ -\frac{\Lambda(5\Lambda a^2-2)ac}{{{(\frac{3}{2}-{k\!\!\!\!^{-} {}})}} }, \ -2{{k\!\!\!\!^{-} {}}}, \ -2i{{%
k\!\!\!\!^{-} {}}}^\frac{1}{2}\Bigr)  \nonumber \\
\overline m^i &=& \overline {P}\Bigl(0, \ -\frac{\Lambda(5\Lambda a^2-2)ac}{{{(\frac{3}{2}-{k\!\!\!\!^{-} {}})}} }, \ -2{{k\!\!\!\!^{-} {}%
}}, \ 2i{{k\!\!\!\!^{-} {}}}^\frac{1}{2}\Bigr)  \nonumber \\
n^i &=& \frac{{Q}}{a}\Bigl( F{{k\!\!\!\!^{-} {}}}^{\frac{1}{4}}, \ L, \ {k\!\!\!\!^{-} {}}^2 c, \ \frac{{k\!\!\!\!^{-} {}}^{\frac{1}{4}}G}{2}\Bigr)
\label{frameH1}
\end{eqnarray}
Since the function $G$ is a solution of the partial table (\ref{sGHPtableG}%
) we can write $-2G=\beta_{3}(t)$, and similarly $-4F=\beta _{2}(t)$; both
are arbitrary functions of $t$, but the latter has the constraint that $%
\beta _{2}(t)\neq 0$.

The metric in $t,c,a,y$ coordinates is therefore given by
\begin{equation}
g^{ij}= \left(
\begin{array}{lccr}
0 & \frac{{k\!\!\!\!^{-} {}}^{1/4}\beta_2(t)}{a{(\frac{3}{2}-{k\!\!\!\!^{-} {}})}%
} & 0 & 0 \cr \frac{{k\!\!\!\!^{-} {}}^{1/4}\beta_2(t)}{ a(\frac{3}{2}-{%
k\!\!\!\!^{-} {}})} & -\frac{8 }{a {{k\!\!\!\!^{-} {}}}^{1/2}(\frac{3}{2}-{%
k\!\!\!\!^{-} {}})^{2}}Z & -\frac{2{{k\!\!\!\!^{-} {}}}(5\Lambda^2 a^4+1)c}{a(\frac{3}{2}-{%
k\!\!\!\!^{-} {}})} &  \frac{{k\!\!\!\!^{-} {}}^{1/4}\beta_3(t)}{a(\frac{3}{2}-{%
k\!\!\!\!^{-} {}})}\cr 0 & -\frac{2{{k\!\!\!\!^{-} {}}}(5\Lambda^2 a^4+1)c}{a(\frac{3%
}{2}-{k\!\!\!\!^{-} {}})} & -4 {{k\!\!\!\!^{-} {}}}^2 & 0 \cr 0 & \frac{{%
k\!\!\!\!^{-} {}}^{1/4}\beta_3(t)}{a(\frac{3}{2}-{k\!\!\!\!^{-} {}})} & 0 &
-4{k\!\!\!\!^{-} {}}%
\end{array}\right)  \label{metric2}
\end{equation}
where
\begin{eqnarray}
Z &=& {(\frac{3}{2}-{k\!\!\!\!^{-} {}})}{{k\!\!\!\!^{-} {}}}^{1/2}\Bigl(L  +\frac{\Lambda^2(5\Lambda a^2-2)^2a^3c^2 }{8
{(\frac{3}{2}-{k\!\!\!\!^{-} {}})} }\Bigr) \nonumber \\
& =& S+{2\sqrt{2}a  }t  -\frac{2 {{{k\!\!\!\!^{-} {}}}^{1/2}} }{\Lambda}
 + \frac {9\Lambda^2 a^3 c^2 {k\!\!\!\!^{-} {}}^{3/2}}{4}
+\frac{{{k\!\!\!\!^{-} {}}}^{1/2}\Lambda^2(5\Lambda a^2-2)^2a^3c^2 }{8
} \nonumber \\
&=&  {\ \frac{2}{\lambda }e^{-\lambda y}} +{2\sqrt{2}a  }t  -\frac{2 {{{k\!\!\!\!^{-} {}}}^{1/2}} }{\Lambda}
 +\frac {\Lambda^2 {{k\!\!\!\!^{-} {}}}^{1/2} a^3 c^2 (25\Lambda^2a^4-2\Lambda a^2+13)} {8}
 \label{Zdefns}
\end{eqnarray}
and ${k\!\!\!\!^{-}{}}={\lambda }^{2}a^{2}+1/2$ from (\ref{A1defn}).

We emphasise that this case only exists for negative $\Lambda=-\lambda^2$.

\subsection{Generic case combined with special case, $V =\hbox{ constant}$}

\label{genne3}

It will be useful to place this  special case (with the cosmetic
 changes $y \to  x$, and $\beta_2(t)\to \alpha_2(t), \beta_3(t)\to \alpha_3(t)$) alongside
the generic metric obtained in the previous section; so we combine the result in the previous
subsection with the generic result in Section \ref{intne} to present the metric in the coordinates $t,c,a,{x}$, given by
\begin{equation}
g^{ij}= \left(
\begin{array}{lccr}
0 & \frac{{k\!\!\!\!^{-} {}}^{1/4}\alpha_2(t)}{a{(\frac{3}{2}-{k\!\!\!\!^{-} {}})}%
} & 0 & 0 \cr \frac{{k\!\!\!\!^{-} {}}^{1/4}\alpha_2(t)}{ a(\frac{3}{2}-{%
k\!\!\!\!^{-} {}})} & -\frac{8 }{a {{k\!\!\!\!^{-} {}}}^{1/2}(\frac{3}{2}-{%
k\!\!\!\!^{-} {}})^{2}}Z & -\frac{2{{k\!\!\!\!^{-} {}}}(5\Lambda^2 a^4+1)c}{a(\frac{3}{2}-{%
k\!\!\!\!^{-} {}})} &  \frac{{k\!\!\!\!^{-} {}}^{1/4}\alpha_3(t)}{a(\frac{3}{2}-{%
k\!\!\!\!^{-} {}})}\cr 0 & -\frac{2{{k\!\!\!\!^{-} {}}}(5\Lambda^2 a^4+1)c}{a(\frac{3%
}{2}-{k\!\!\!\!^{-} {}})} & -4 {{k\!\!\!\!^{-} {}}}^2 & 0 \cr 0 & \frac{{%
k\!\!\!\!^{-} {}}^{1/4}\alpha_3(t)}{a(\frac{3}{2}-{k\!\!\!\!^{-} {}})} & 0 &
-4{k\!\!\!\!^{-} {}}%
\end{array}\right)  \label{metric5}
\end{equation}
where $\alpha _{3}(t)$ is an arbitrary function of $t$ including the zero
function, whereas $\alpha _{2}(t)$ is an arbitrary function of $t$ excluding
the zero function, and ${k\!\!\!\!^{-}{}}=\Lambda a^{2}+1/2$.

There are two
possibilities for $Z$:
\begin{eqnarray}
\hbox{(i)}\
Z =   \alpha_1(t)\hbox{cos[h]}(\sqrt{|\Lambda|}x) +{2\sqrt{2}a  }t  -\frac{2 {{{k\!\!\!\!^{-} {}}}^{1/2}} }{\Lambda}
\qquad\qquad\qquad\qquad \nonumber \\ +\frac {\Lambda^2 {{k\!\!\!\!^{-} {}}}^{1/2} a^3 c^2 (25\Lambda^2a^4-2\Lambda a^2+13)} {8}\qquad   \label
{Zcoords4}
\end{eqnarray}
from (\ref{Zcoords1})  where $\alpha _{1}(t)\ne 0$ is an
arbitrary function of $t$ excluding the zero function, and $\hbox{cos[h]}(\sqrt{|\Lambda|}x) $ is given by (\ref{cosh}).

\begin{eqnarray}
\hbox{(ii)}\ Z =  {\ \frac{2}{\lambda }e^{-\lambda x}} +{2\sqrt{2}a  }t  -\frac{2 {{{k\!\!\!\!^{-} {}}}^{1/2}} }{\Lambda}
 +\frac {\Lambda^2 {{k\!\!\!\!^{-} {}}}^{1/2} a^3 c^2 (25\Lambda^2a^4-2\Lambda a^2+13)} {8}  \label{Zcoords5}
\end{eqnarray}
from (\ref{Zdefns}).

Note that case (i) exists for positive and negative cosmological constant,
but case (ii) only exists for negative $\Lambda$, with $\lambda=\pm \sqrt{%
-\Lambda}$.

\section{The most general form for the metric when $\Lambda + \protect\tau%
\overline \protect\tau \ne 0$.}

\label{all}

\subsection{Preliminaries to generalisations}\label{all;1}

We have not yet got the most general version of the metric because in
Section \ref{intne} we assumed that $T$ was not a constant in order to be
able to choose it as a coordinate candidate, and we also assumed that $S\ne 0$  in order to be
able to choose $X$ as a coordinate candidate.

We begin with  the excluded case where $T$ is a constant. In such a situation, clearly $%
F=0$ so we cannot instead use $F$  as a coordinate candidate, but we still
have the possibility of choosing $H$ or $S$ as a coordinate candidate. Once
we make such a choice then we could continue in a similar manner as in the
last section, building our tables, and hence the tetrad, around our four
coordinate candidates. However, if \textit{neither} of the other functions $%
H, S$ is functionally independent of the original three coordinates, then it
will \textit{not} be possible to find a replacement candidate \textit{%
directly}; we emphasise that in such circumstances no additional independent
quantities can be generated by any {\it direct} manipulations of the tables and
the commutators. In such a situation we still need a replacement candidate
in order to extract the remaining information from the commutators. So
rather than treating the special case $F= 0$ separately, we will extend the
generic result to include this special case as well.

We shall now show, instead, that a {\it complementary coordinate
candidate} to replace $T$ can easily be found, and then, using this
coordinate, we will obtain a generalisation of the metric (\ref{metric5})
which includes all possible values for $T$, including a constant.

Secondly we consider  the excluded case  $S=0$, and for this case we  find that not only can we not construct  $X$ (or $V$) as a coordinate candidate, but that we cannot generate {\it directly} any replacement coordinate candidate. We shall now show, instead, that a {\it complementary coordinate
candidate} to replace $X$ can easily be found, and then, using this
coordinate, we will first obtain this excluded case $S=0$ separately; we will then obtain a generalisation of the metric (\ref{metric5})
which includes this additional special case,  $S=0$.

\subsection{Finding a complementary coordinate candidate to replace $T$}\label{all;2}

The results in Section \ref{intne} up to the end of subsection \ref{intne4}
apply as before; the only difference here is that we \textit{interpret} them
differently. When we are interpreting our tables and choosing our explicit
coordinate candidates we will now consider only the three zero-weighted real
scalars $A, C, X$ as coordinate candidates while the zero-weighted scalar $T$
is not now included as a coordinate candidate, and so there is now no hindrance
to it acquiring  a constant value, even zero. A related change is
that since $T$ is no longer a coordinate candidate, we no longer need its
{\it complete} table, nor the resulting partial table for $F$; however we still
need the \textit{partial} table for $T$ since it is a result of applying the
commutators to $\mathbf{I}$, and so is still a crucial component of the analysis,
\begin{eqnarray}
\hbox{\rm I}\kern-0.32em\raise0.35ex\hbox{\it o} T & = & 0  \nonumber \\
\hbox{$\partial$\kern-0.25em\raise0.6ex\hbox{\rm\char'40}} T & = & 0
\nonumber \\
\hbox{$\partial$\kern-0.25em\raise0.6ex\hbox{\rm\char'40}}^{\prime}T & = & 0
\label{partialtableT1}
\end{eqnarray}

So, clearly we do not have our full quota of \textit{four} coordinate
candidates, but we do not wish to use any of the remaining intrinsic quantities from
the tables, since it would involve the additional assumption of that
quantity being non-constant.

It is now very important to note that all the \textit{direct} information
which can be obtained from the intrinsic elements of the GHP formalism is in
these tables, and no amount of further manipulation  of the equations with the
commutators will generate a replacement coordinate candidate which is
functionally independent of the other three $A, C, X$. On the otherhand, we
have not yet extracted all the information from the commutators since we
have only applied them to \textit{three} zero-weighted scalars. So we
require a fourth zero-weighted scalar --- functionally independent of the
other three $A, C, X$ --- which will be the fourth coordinate candidate, and
also enable us to extract any remaining information implicit in the
commutators. Since there is no such intrinsic zero-weighted scalar which we
can generate directly in the GHP formalism, we introduce it \textit{%
indirectly} via its table, which will have to be consistent with all the
explicit equations in the GHP formalism, and in particular with the GHP
commutators

 In fact, we get a strong hint from Section \ref{intne4}, by
looking at the table (\ref{GHPtableT}) for the coordinate $T$ (which is the
missing coordinate candidate in this case); so we consider the possibility
of the existence of a real zero-weighted scalar $\tilde T $, which satisfies
the table\footnote{%
For easy reference, in an extended case, we will label by $\tilde{T}$ a
\textit{complementary} coordinate candidate which replaces an intrinsic coordinate
candidate $T$ in the corresponding  generic case; but we emphasise this is not to imply any
{\it direct} link between the two quantities, it simply points us to the source of
the hint which suggested the table for the complementary coordinate
candidate.}
\begin{eqnarray}
\hbox{\rm I}\kern-0.32em\raise0.35ex\hbox{\it o} \tilde T & = & 0  \nonumber
\\
\hbox{$\partial$\kern-0.25em\raise0.6ex\hbox{\rm\char'40}} \tilde T & = & 0
\nonumber \\
\hbox{$\partial$\kern-0.25em\raise0.6ex\hbox{\rm\char'40}}^{\prime}\tilde T
& = & 0  \nonumber \\
\hbox{\rm I}\kern-0.32em\raise0.35ex\hbox{\it o}^{\prime}\tilde T & = & {-4
\frac{{Q}}{A}} {{k\!\!\!\!^{-} {}}}^{ \frac{1}{4}}
\label{GHPtabletildeT}
\end{eqnarray}
(A direct copy of (\ref{GHPtableT}) would suggest the table
\begin{eqnarray}
\hbox{\rm I}\kern-0.32em\raise0.35ex\hbox{\it o} \tilde T & = & 0  \nonumber
\\
\hbox{$\partial$\kern-0.25em\raise0.6ex\hbox{\rm\char'40}} \tilde T & = & 0
\nonumber \\
\hbox{$\partial$\kern-0.25em\raise0.6ex\hbox{\rm\char'40}}^{\prime}\tilde T
& = & 0  \nonumber \\
\hbox{\rm I}\kern-0.32em\raise0.35ex\hbox{\it o}^{\prime}\tilde T & = & {%
\frac{{Q}F(\tilde T)}{A}} {{k\!\!\!\!^{-} {}}}^{ \frac{1}{4}}
\label{GHPtabletildeT2}
\end{eqnarray}
where $F(\tilde T)$ is an arbitrary function of $%
\tilde T$, excluding the zero function. However it is easy to see that a
simple coordinate transformation $\tilde T \to -4\int F(\tilde T) d \tilde T$ gives the simpler version (\ref%
{GHPtabletildeT}).)

So we have chosen a zero-weighted real scalar $\tilde T$ defined by its table (\ref%
{GHPtabletildeT}), whose structure we have 'copied' from the table structure (%
\ref{GHPtableT}) of $T$.

It is important to appreciate the different natures of $T$ and $\tilde T$. In Section \ref{intne}, $T$ was defined directly in terms of intrinsic elements
of the formalism, and so was itself an \textit{intrinsic} coordinate
candidate, and the table (\ref{GHPtableT}) was a consequence of its
definition; on the otherhand, the \textit{complementary} coordinate
candidate $\tilde T$ is not defined in terms of intrinsic quantities of the
formalism, but rather as the integral of the table (\ref{GHPtabletildeT}). Hence, the introduction
of the coordinate candidate $\tilde T$, via the table (\ref{GHPtabletildeT}%
), is structurally different from the usual direct identification of
coordinates with elements of the formalism: $C,A,X$ are intrinsic coordinate
candidates, while $\tilde T$ is a complementary coordinate candidate.

It is straightforward to confirm that this choice of table (\ref{GHPtabletildeT}%
) is consistent with the GHP commutators (\ref{GHPcomm}) and creates no
inconsistency with the other tables.

So, compared to Section \ref{intne}, we have simply replaced the fourth
\textit{intrinsic} coordinate candidate $T$ with the \textit{complementary}
coordinate candidate $\tilde T$ defined via its table (\ref{GHPtabletildeT})
whose structure was 'copied' from the table (\ref{GHPtableT}) for $T$; in addition we remember that
the real zero-weighted quantity $T$ now satisfies (\ref{partialtableT1}). Clearly $T$ now is a function of only the one coordinate candidate $\tilde T$, i.e., $T(\tilde T)$. The
remaining tables are unchanged.

\subsection{Finding a complementary coordinate candidate to replace $X$}\label{all;3}

The results in Section \ref{intne} up to the end of subsection \ref{intne3}
apply as before; and we shall also assume the results up to equation (\ref{Sdefn}).

When we make the substitution $S=0$ into (\ref{GHPtableS}) we find that the table collapses giving $B=0$.  This means that the table for $B$,  (\ref{GHPtableB}) also collapses.  No action with the commutators is able to generate
any new information {\it directly} from the existing GHP equations. At this stage we are left with only the GHP tables for the three coordinate candidates $A,C,T$ and the GHP tables for the weighted scalars, $P,Q$. However we need a table for a fourth coordinate candidate in order to be able to extract all the information from the GHP commutators.
 So we
require a fourth zero-weighted scalar --- functionally independent of the
other three $A, C, T$ --- which will be the fourth coordinate candidate, and
also enable us to extract any remaining information implicit in the
commutators. So, in a similar manner to the last subsection, we introduce a {\it complementary} coordinate candidate \textit{indirectly} via its table, which will have to be consistent with all the
explicit equations in the GHP formalism, and in particular with the GHP
commutators.

Also, as in last section,  we get a strong hint from Section \ref{intne4}, by
looking at the table (\ref{GHPtableX}) for the coordinate $X$ (which is the
missing coordinate candidate in this case); so we consider the possibility
of the existence of a real zero-weighted scalar $\tilde X $, which satisfies
the table
\begin{eqnarray}
\hbox{\rm I}\kern-0.32em\raise0.35ex\hbox{\it o} {\tilde X} &=& 0  \nonumber \\
\hbox{$\partial$\kern-0.25em\raise0.6ex\hbox{\rm\char'40}} {\tilde X} &=& -2i%
P{{k\!\!\!\!^{-} {}}}^\frac{1}{2}  \nonumber \\
\hbox{$\partial$\kern-0.25em\raise0.6ex\hbox{\rm\char'40}}^{\prime}{\tilde X} &=&2i%
\overline {P}{{k\!\!\!\!^{-} {}}}^ \frac{1}{2}  \nonumber \\
\hbox{\rm I}\kern-0.32em\raise0.35ex\hbox{\it o}^{\prime}{\tilde X} &=& \frac{{%
Q}{{k\!\!\!\!^{-} {}}}^{ \frac{1}{4}}}{2A} H \quad
\label{GHPtabletildeX}
\end{eqnarray}
where we also assume $H(t)$.

Again we have adopted the convention of  labelling by $\tilde{X}$ a
\textit{complementary} coordinate candidate which replaces an intrinsic coordinate
candidate $X$ in the corresponding  generic case.

It is straightforward to confirm that this choice of table (\ref{GHPtabletildeX}) is consistent with the GHP commutators (\ref{GHPcomm}) and creates no
inconsistency with the other tables.

Furthermore, we note  since $\tilde{X}$ is a complementary coordinate candidate which does not occur except in its own table, that we could have made an even  simpler choice of table, by choosing $H=0$ (which can easily be confirmed by a coordinate transformation $\tilde X \to \tilde X + \int  (H(t)/4) dt$.) However, we shall not make that simplification, for presentation reasons.

We can therefore present this special case in the coordinates $t,c,a,{\tilde x}$, as
\begin{equation}
g^{ij}= \left(
\begin{array}{lccr}
0 & \frac{{k\!\!\!\!^{-} {}}^{1/4}\alpha_2(t)}{a{(\frac{3}{2}-{k\!\!\!\!^{-} {}})}%
} & 0 & 0 \cr \frac{{k\!\!\!\!^{-} {}}^{1/4}\alpha_2(t)}{ a(\frac{3}{2}-{%
k\!\!\!\!^{-} {}})} & -\frac{8 }{a {{k\!\!\!\!^{-} {}}}^{1/2}(\frac{3}{2}-{%
k\!\!\!\!^{-} {}})^{2}}Z & -\frac{2{{k\!\!\!\!^{-} {}}}(5\Lambda^2 a^4+1)c}{a(\frac{3}{2}-{%
k\!\!\!\!^{-} {}})} &  \frac{{k\!\!\!\!^{-} {}}^{1/4}\alpha_3(t)}{a(\frac{3}{2}-{%
k\!\!\!\!^{-} {}})}\cr 0 & -\frac{2{{k\!\!\!\!^{-} {}}}(5\Lambda^2 a^4+1)c}{a(\frac{3%
}{2}-{k\!\!\!\!^{-} {}})} & -4 {{k\!\!\!\!^{-} {}}}^2 & 0 \cr 0 & \frac{{%
k\!\!\!\!^{-} {}}^{1/4}\alpha_3(t)}{a(\frac{3}{2}-{k\!\!\!\!^{-} {}})} & 0 &
-4{k\!\!\!\!^{-} {}}%
\end{array}
\right)  \label{metric1.1}
\end{equation}
where ${k\!\!\!\!^{-} {}}=\Lambda a^2 +1/2$, and $Z$ is given  by,
\begin{eqnarray}
Z ={2\sqrt{2}a  }t  -\frac{2 {{{k\!\!\!\!^{-} {}}}^{1/2}} }{\Lambda}
+\frac {\Lambda^2 {{k\!\!\!\!^{-} {}}}^{1/2} a^3 c^2 (25\Lambda^2a^4-2\Lambda a^2+13)} {8}
  \label{Zcoords1.1}
\end{eqnarray}
and   $\alpha _{3}(t)$ is completely arbitrary, while $\alpha _{2}(t)$ is arbitrary, except for the zero function.
\smallskip

It is clear that this special case simply fills the gap in our original case (\ref{metric1}), (\ref{Zcoords1}) by now including the case $\alpha _{1}(t)=0$ which was excluded there.

\subsection{The most general metric}\label{all;4}

The metric (\ref{metric5}) gives the most general form of the metric for this class of spaces --- under the additional restrictions that no Killing vectors are present. This follows from the existence of {\it four intrinsic coordinates}; this is also  confirmed in \cite{edram4}  where we consider the detailed invariant Karlhede classification of this class of metrics.   In subsection \ref{all;2}  we saw how to generalise (\ref{metric5}) to  include the possibility of the coordinate $\tilde t$ being a complementary coordinate, so that this more general class also permits the existence of a Killing vector. The special case (\ref{metric1.1}) just deduced in subsection \ref{all;3} can also easily be generalised in the same manner  by replacing $t$ with a complementary coordinate $\tilde t$;  this special case could then be listed alongside  the generalisation of (\ref{metric5}). However it is more convenient to simply incorporate (\ref{metric1.1}) into
the generalisation of (\ref{metric5}) discussed in subsection \ref{all;2} , by just removing the restriction $\alpha_1(t) \ne 0$. It is easy to confirm that  the tables for the respective complementary candidates $\tilde T$ and $\tilde  X$ are consistent with all the other tables, and with each other.

Hence we  generalise the combined metric form (\ref{metric5})
given in the last section by replacing the intrinsic coordinate candidate $T$ and its
table with the complementary coordinate $\tilde T$ and its table, and the intrinsic coordinate candidate $X$ and its
table with the complementary coordinate $\tilde X$ and its table, and finally   obtaining  the metric in
the coordinates
\[
\tilde t= \tilde T,\qquad c= C, \qquad a=A, \qquad  \tilde x=  \tilde X \ , %
\label{coordsomenumber1}
\]
given by
\begin{equation}
g^{ij}= \left(
\begin{array}{lccr}
0 & \frac{{k\!\!\!\!^{-} {}}^{1/4}}{a{(\frac{3}{2}-{k\!\!\!\!^{-} {}})}%
} & 0 & 0 \cr \frac{{k\!\!\!\!^{-} {}}^{1/4}}{ a(\frac{3}{2}-{%
k\!\!\!\!^{-} {}})} & -\frac{8 }{a {{k\!\!\!\!^{-} {}}}^{1/2}(\frac{3}{2}-{%
k\!\!\!\!^{-} {}})^{2}}Z & -\frac{2{{k\!\!\!\!^{-} {}}}(5\Lambda^2 a^4+1)c}{a(\frac{3}{2}-{%
k\!\!\!\!^{-} {}})} &  \frac{{k\!\!\!\!^{-} {}}^{1/4}\gamma_3(\tilde t)}{a(\frac{3}{2}-{%
k\!\!\!\!^{-} {}})}\cr 0 & -\frac{2{{k\!\!\!\!^{-} {}}}(5\Lambda^2 a^4+1)c}{a(\frac{3%
}{2}-{k\!\!\!\!^{-} {}})} & -4 {{k\!\!\!\!^{-} {}}}^2 & 0 \cr 0 & \frac{{%
k\!\!\!\!^{-} {}}^{1/4}\gamma_3(\tilde t)}{a(\frac{3}{2}-{k\!\!\!\!^{-} {}})} & 0 &
-4{k\!\!\!\!^{-} {}}%
\end{array}\right)  \label{metric6}
\end{equation}
where $\gamma_3(\tilde t)$ is an arbitrary function of $\tilde t$ including
the zero function, and ${k\!\!\!\!^{-} {}}=\Lambda a^2+1/2$. There are two
possibilities for $Z$,
\begin{eqnarray}
\hbox{(i)}\    Z =  \gamma_1(\tilde t)\hbox{cos[h]}(\sqrt{|\Lambda|}\tilde x) &+&{2\sqrt{2}a  }\gamma_2(\tilde t)  -\frac{2 {{{k\!\!\!\!^{-} {}}}^{1/2}} }{\Lambda} \qquad\qquad\qquad\qquad \qquad\qquad
\nonumber \\ &+&\frac {\Lambda^2 {{k\!\!\!\!^{-} {}}}^{1/2} a^3 c^2 (25\Lambda^2a^4-2\Lambda a^2+13)} {8}
\label{L*coords61}
\end{eqnarray}
where $\gamma_1(\tilde t)$ and $\gamma_2(\tilde t)$ are arbitrary functions
of $\tilde t$ including the zero function.
\begin{eqnarray}
\hbox{(ii)}\ Z =  {\ \frac{2}{\lambda }e^{-\lambda \tilde x} +{2\sqrt{2}a  } \gamma_2(\tilde t) -\frac{2 {{{k\!\!\!\!^{-} {}}}^{1/2}} }{\Lambda}
 +\frac {\Lambda^2 {{k\!\!\!\!^{-} {}}}^{1/2} a^3 c^2 (25\Lambda^2a^4-2\Lambda a^2+13)} {8}}
 \label{L*coords72}
\end{eqnarray}
where $\gamma_2(\tilde t)$ is an arbitrary function of $\tilde t$ including
the zero function. The changes $\alpha_1(t) \to \gamma_1(t),\  \alpha_2(t) \to \gamma_2(t),\ \alpha_3(t) \to \gamma_3(t)$ are simply cosmetic.

Note that case (i) exists for positive and negative cosmological constant,
but case (ii) only exists for negative $\Lambda$, with $\lambda=\pm \sqrt{%
-\Lambda}$.

When we compare the metric (\ref{metric5}) where $Z$ is given by (\ref%
{Zcoords4}) or (\ref{Zcoords5}) with the above metric (\ref{metric6}) where $%
Z$ is given by (\ref{L*coords61}) or (\ref{L*coords72}), we can easily
demonstrate that the former is a special case of the latter, by making the
coordinate transformation $t=\gamma_2(\tilde t)/2\sqrt{2}$, and identifying $%
\gamma_1(\tilde t) = \gamma_1\bigl(\gamma_2^{-1}(2\sqrt{2}t)\bigr) =
\alpha_1(t)$ and $\gamma_3(\tilde t) = \gamma_3\bigl(\gamma_2^{-1}(2\sqrt{2}%
t)\bigr) = \alpha_3(t)$, we confirm that the former case is included in the
latter. However the latter also permits $\gamma_2(\tilde t)$ to be constant,
even zero; this is a possibility missing from the former.

It is trivial to confirm that the special  subclass (\ref{metric1.1}) is simply the special case of (i) given by $\gamma_1(\tilde t)=0$. We note that we have used the notation $\tilde x$ in this general form, although it is obvious that this coordinate is in fact an intrinsic coordinate --- except in this very special case $\gamma_1(\tilde t)=0$.
Finally, we  note again that
in this very special case  $\gamma_1(\tilde t)=0$ a simple
coordinate transformation gives $\gamma_3(\tilde t)=0$,  but leaves everything
else unchanged.

\section{Summary and Discussion}

The class of conformally flat pure radiation spacetimes with a non-zero cosmological
constant which have been studied in \cite{edram2} and in this paper has
provided a very good laboratory for developing techniques and increasing our
experience in the GIF formalism.
We have shown how the method in \cite{edvic} which was used to investigate
conformally flat pure radiation spacetimes (with $\tau \ne 0$) can be developed to investigate
the more complicated situation where, in addition, there is a non-zero
cosmological constant; in \cite{edram2} we have found the subclass of
conformally flat pure radiation spacetimes with negative cosmological
constant $\Lambda = - \tau\bar\tau ,  \tau \ne 0$, while in this paper we have found the remaining subclass with $\Lambda + \tau\bar\tau \ne 0 \ne \tau$.

An important new development in this paper is the realisation that we do not need to work the whole integration procedure in the GIF, but rather we can change to the GHP formalism once the GIF has generated the second spinor and  we have extracted  information by applying the GIF commutators to this spinor; since calculations in the GIF can be long and complicated, it is a considerable advantage to be able to transfer to the GHP formalism for the bulk of the calculations and only use the GIF in the initial calculations associated with the determination of  the second spinor. In this paper, the  simplification of the tables in subsection 5.4,  which was crucial in order to obtain such a manageable form for the eventual  metric, would not have been so transparent and would have been much more complicated in GIF.

 This integration procedure within the GIF/GHP formalism is particularly suited to spaces with four intrinsic coordinates; spaces with less than four intrinsic coordinates may appear to pose more difficulties. Another important development in this paper is a fuller understanding of how `generic' results help to suggest additional special cases; in the case where it is suspected that there exists additional special cases to the generic case, the structure of tables for complementary coordinates can be `copied' from the corresponding intrinsic coordinates.

In addition,  in \cite{edram2} we learned how to
treat the one dimensional isotropy freedom of a null rotation. These various calculations and results
are enabling us gradually to build up our experience and skill in the GIF/GHP formalism,
with the ultimate goal of tackling even more complicated situations in the
future.The actual metrics which we have obtained  have been confirmed with Maple.

It is clear from the most general form of the metric, and the fact that it is --- as much as possible --- presented in essential coordinates, that there  will be subclasses with zero, one and two Killing vectors. There is in fact a rich symmetry structure in the whole class of conformally flat pure radiation  spacetimes with non-zero cosmologicak constant, and the full details  are presented in \cite{edram4}.

As well as increasing our experience and expertise in the GIF operator
integration method, this particular class of spaces is interesting in its
own right. The analogous spacetimes with zero cosmological constant  investigated in \cite{edvic}  revealed some complications and subtleties in the computer classification programmes \cite{classi}, \cite{GR}; it will be interesting to see how the computer programmes handle these new spacetimes, and especially the existence of one degree of null isotropy.  It will also be interesting to explore the physical interpretation of the spacetimes in this paper and in \cite{edram2}, along the lines investigated  in \cite{gr1}  for the spaces  with zero cosmological constant;  the wide variety of individual
subclasses with a range from zero to five Killing vectors give a rich area of investigation.

 In some classes
of spacetimes the addition of a cosmological constant makes little
significant difference. On
the otherhand, the addition of a non-zero cosmological constant has made a
significant difference to vacuum Petrov type D spaces \cite{czmc}, \cite%
{carm1}, \cite{carm2}. Moreover, its addition to the non-expanding Kundt
class of spacetimes significantly complicates the equations: some classes of
Petrov type N with non-zero cosmological constant have been found and analysed \cite{ozs}, \cite{bic1}, \cite%
{bic2}, \cite{pod} as well as some of Petrov type II \cite{pod}; recently a
detailed and comprehensive derivation and analysis of Petrov type III
non-expanding vacuum spacetimes with non-zero cosmological constant has been
carried out in \cite{gr1}.

It may be suspected that these various examples of Type II, III and N spaces
just mentioned will specialise in the conformally flat limit to the spaces
under consideration in this paper. However, that is not necessarily so,
since, in at least some of those investigations, properties of a non-zero
Weyl tensor were built into the analysis. Furthermore, even if the
conformally flat limit does exist in some of the investigations, the form of
the metric may  be much more complicated than in our version where we have built the
structure  around the conformally flat properties from the
beginning. It has therefore been of interest to see how our method supplies 'good' coordinates, simple
differential equations, and a very manageable form for the metrics.
However, a number of terms have square roots, as well as trignometric and hyperbolic functions, and  absolute value functions have been used in the calculations; these will put restrictions on the range of the coordinates, and there will be alternative, and possibly more general, coordinate systems to consider. It remains to investigate the whole class of
these spacetimes found via GIF, considering in more detail the coordinate systems, and comparing with the conformally flat limits of these
various other investigations.

\

\

{\bf Acknowledgements}

SBE wishes to thank Officina Mathematica for supporting a visit to
Universidade do Minho and the Department of Mathematics for Science
and Technology for their hospitality. MPMR wishes to thank
Vetenskapsr\aa det (Swedish Research Council) for supporting a visit
to Link\"opings universitet and the Mathematics Department for their
hospitality.

\end{document}